\documentclass[11pt]{article}
\pdfoutput=1
\usepackage{jcappub,natbib}
\usepackage{graphicx}
\usepackage{bbm}
\usepackage{amsmath}
\usepackage{amssymb}
\usepackage{braket}
\usepackage{color} 
\input{colordvi.tex}

\def\ga{\mathrel{\raise.3ex\hbox{$>$\kern-.75em\lower1ex\hbox{$\sim$}}}}
\def\la{\mathrel{\raise.3ex\hbox{$<$\kern-.75em\lower1ex\hbox{$\sim$}}}}

\def\bk{{\bf k}}
\def\bq{{\bf q}}

\def\bp{{\bf p}}

\def\d{\delta}

\def\bk{{\bf k}}
\def\bq{{\bf q}}

\def\bp{{\bf p}}
\def\la{~\mbox{\raisebox{-.6ex}{$\stackrel{<}{\sim}$}}~}
\def\ga{~\mbox{\raisebox{-.6ex}{$\stackrel{>}{\sim}$}}~}

\title{Generation of gravitational waves from symmetry restoration during inflation}

\author{Cody Goolsby-Cole and Lorenzo Sorbo} 

\affiliation{Amherst Center for Fundamental Interactions, Department of Physics, University of Massachusetts, Amherst, MA 01003}

\abstract{We discuss the possibility of a feature in the spectrum of inflationary gravitational waves sourced by a scalar field $\chi$ whose vacuum fluctuations are amplified by a rapidly time dependent mass.  Unlike previous work which has focused on the case in which the mass of the field $\chi$ vanishes only for an instant before becoming massive again, we study a system where the scalar field becomes and remains massless through the end of inflation as the consequence of the restoration of a shift symmetry.  After applying appropriate constraints to our parameters, we find, for future CMB experiments, a small contribution to the tensor-to-scalar ratio which can be at most of the order $r \sim 10^{-5}$. At smaller scales probed by gravitational interferometers, on the other hand, the energy density in the gravitational waves produced this way might be above the projected sensitivity of LISA, $\Omega_{GW}\,h^2 \sim 10^{-13}$, in a narrow region of parameter space.  If there is more than one $\chi$ species, then these amplitudes are enhanced by a factor equal to the number of those species. }

\begin{document}

\begin{flushright}  ACFI-T17-11  \end{flushright}

\maketitle
\flushbottom


\section{Introduction}%

Inflation generates a isotropic, homogeneous, flat Universe with a spectrum of scalar perturbations. On the top of this, inflation also produces a spectrum of primordial gravitational waves (PGWs) -- see~\cite{Guzzetti:2016mkm} for a recent review.  The contribution to the graviton power spectrum produced by a pure de Sitter expansion with constant expansion rate $H$ is~\cite{Starobinsky:1979ty}
\begin{equation} \label{psnom}
\mathcal{P}_T^{\rm vacuum}= \frac{2}{\pi^2} \frac{H^2}{M_P^2} \,.
\end{equation}
The above relation provides a simple and yet powerful prediction of inflation, which allows us to connect the energy scale of inflation to an observable quantity.

While the stochastic background amplitude~(\ref{psnom}) of the spectrum of PGWs generated during inflation is model dependent and might be too small to be observable,  the detection of PGWs through the Cosmic Microwave Background would certainly represent a major result in support of inflation.    The current upper bound on tensor modes produced during inflation for a single field model is provided by the BICEP/Keck collaboration, that, after including other constraints from cosmological measurements, finds the limit $r<.07$~\cite{Array:2015xqh}, where $r$ is the tensor-to-scalar ratio defined as $r\equiv\mathcal{P}_T/\mathcal{P}_\zeta\simeq 4.5\times 10^8\,\mathcal{P}_{T} $.  Future CMB experiments aim at pushing this limit further. In particular,  the next generation CMB-S4 experiment aims at a tensor-to-scalar ratio sensitivity of $r \sim 10^{-4}$~\cite{Abazajian:2016yjj}.  {\em Direct} detection,  in the near future, of the stochastic PGW background generated during inflation from amplification of vacuum fluctuations is unlikely due to CMB constraints~\cite{Array:2015xqh} which yield an upper bound $\Omega_{GW}\, h^2 \lesssim 10^{-15}$ on the energy density of PGWs. Far future experiments such as BBO or DECIGO, however, aim at sensitivies of the order $\Omega_{GW}\, h^2 \sim 10^{-15} - 10^{-17}$ \cite{Corbin:2005ny,Kudoh:2005as}.

There has been an increasing interest in the possibility of disentangling the value of $r$ from the energy scale of inflation by adding new sources of tensor modes. Such an interest was partly motivated by an early belief~\cite{Kallosh:2007wm} that models of inflation in String Theory generally take place at such low energies that $r$ is small and unobservable.  Moreover, alternative mechanisms producing gravitational waves lead in general to a phenomenology that is much richer than that of the ``standard'' PGWs generated by the amplification of vacuum fluctuations, which have a featureless, slightly red power spectrum, do not violate parity, and do not present any detectable nongaussianities. In particular, models where the inflaton is coupled to gauge fields through a parity-violating interaction have been shown to be able to generate a spectrum of PGWs where all those properties of vacuum tensors are violated to some degree~\cite{Sorbo:2011rz, Barnaby:2011qe,Anber:2012du,Barnaby:2012xt,Crowder:2012ik,Cook:2013xea,Mukohyama:2014gba,Caprini:2014mja,Namba:2015gja,Domcke:2016bkh,Shiraishi:2016yun,Obata:2016xcr,Dimastrogiovanni:2016fuu,Adshead:2016omu,Fujita:2017jwq,Adshead:2017hnc}. Reference~\cite{Anber:2016yqr} has considered the case where chiral fermions are sourcing PGWs. The possibility that the PGW spectrum shows some features implies, in particular, that those PGWs might even be directly detectable by interferometers, as first proposed in~\cite{Cook:2011hg} and also discussed in~\cite{Barnaby:2011qe,Barnaby:2012xt,Crowder:2012ik,Obata:2016xcr} (the work~\cite{Bartolo:2016ami} refers to much of the literature on this topic). 

Models generating additional tensor modes usually assume the existence of a sector whose finite momentum modes are for some reason excited during inflation and act as a classical source of tensors~\cite{Cook:2011hg,Senatore:2011sp}. One the simplest and most studied systems where a sector gets excited during inflation is that of a scalar field $\chi$ that interacts with the inflaton $\phi$ through the coupling~\cite{Chung:1999ve,Barnaby:2009mc}
\begin{align}\label{oldchiphi}
{\cal L}_{\phi\chi}=-\frac{g^2}{2}\left(\phi-\phi_*\right)^2\,\chi^2\,,
\end{align}
with $\phi_*$ a constant. If, as is the case during inflation, the spatial gradients of $\phi$ are negligible, the coupling~(\ref{oldchiphi}) can be seen as  an effective mass $m_\chi=g\,\left|\phi-\phi_*\right|$ for $\chi$. When $m_\chi$ crosses zero (that is, when $\phi$ crosses $\phi_*$), quanta of $\chi$ with momenta up to $\sim\sqrt{g\,|\dot\phi|}$ are excited~\cite{Dolgov:1989us,Traschen:1990sw,Kofman:1997yn}. Those quanta act in their turn as a source of gravitational waves, whose amplitude was first computed in \cite{Cook:2011hg,Senatore:2011sp} and was found not to be competitive with that of the PGWs generated by the amplification of vacuum fluctuations, eq.~(\ref{psnom}). More specifically, by choosing the coupling $g=1$ to maximize the effect, reference~\cite{Cook:2011hg} found that the tensor-to-scalar ratio $r_{\rm sourced}$ of the induced tensors was satisfying the condition 
\begin{align}
\frac{r_{\rm sourced}}{r_{\rm vacuum}}\lesssim 5\times 10^{-7}\left(\frac{r_{\rm vacuum}}{.07}\right)\,,
\end{align}
which was leading to a small and unobservable $r_{\rm sourced}\lesssim 10^{-8}$ even for the largest allowed ${r_{\rm vacuum}}\simeq {.07}$.

The fact that the coupling~(\ref{oldchiphi}) does not induce a sufficiently large amplitude of gravitational waves was interpreted~\cite{Barnaby:2012xt} as a consequence of the fact that, after  crossing $0$, the value of $m_\chi$ obtained from eq.~(\ref{oldchiphi}) starts growing again, rapidly turning the excited modes of $\chi$ into nonrelativisic ones, which are a very inefficient source of gravitational waves. Based on this observation, in the present paper we will consider gravitational waves produced by a scalar $\chi$ that becomes massless during inflation through its coupling to a secondary field $\sigma$ and stays massless afterwards (reference~\cite{Senatore:2011sp} studied, in a construction different from ours, the case where the mass $m_\chi$ converged to a constant after the event of particle creation).  In our model the mass of the field $\chi$ is controlled by a third field $\sigma$ that undergoes symmetry restoration as a consequence of the dynamics of the inflaton. The mass of $\chi$ linearly decreases during its early evolution parameterized by a mass term $-\Lambda_\chi^3(t-t_*)$, and then becomes massless from the time $t_*$ through the end of inflation.

After subtracting unphysical divergences and applying appropriate constraints from CMB observations, we find that for a single scalar field $\chi$ the value of $r_{\text{sourced}}$ is subdominant with respect to the vacuum contribution $r_{\rm vacuum}$ and can be at most of the order of $\sim 10^{-5}$.  This value can be boosted by a factor $N_\chi$ equal to the number of $\chi$ species. 

On the other hand, at the smaller scales probed by interferometers, where we can ignore the constraints that originate from CMB observations, we find an absolute upper bound on the energy density of gravitational waves of $\Omega_{GW}\,h^2 \lesssim 10^{-12}$ (which again can be enhanced by a factor $N_\chi$) which is obtained by saturating a number of inequalities. For ``natural'' choices of parameters, however, we expect to find values of $\Omega_{GW}\,h^2$ are a few orders of magnitude smaller. For comparison, amplitudes of the order of $\Omega_{GW}\,h^2 \sim 10^{-13}$ would be detectable by LISA~\cite{Bartolo:2016ami}.

Our paper is organized as follows.  In Section~\ref{sec:setup}, we discuss the model for both scalars $\chi$ and the spectator field $\sigma$ as well as the equations for the gravitational waves. In Section~\ref{sec:gwprod}, we calculate the evolution of $\chi$ and $\sigma$ in the de Sitter background and $\chi$'s contribution to the graviton's power spectrum while also discussing our renormalization procedure for removing the divergences that are introduced when $\chi$ becomes massless. In Section~\ref{sec:constraints}, we constrain the parameters of our model by imposing both observational limits and perturbativity requirements. In Section~\ref{sec:gwlarge}, we discuss the maximal amplitude that might be attained by the PGWs produced by our mechanism.  Finally in Section~\ref{sec:conclusion}, we summarize our results and discuss possible future work.

\section{Set Up}
\label{sec:setup}

We examine graviton production in a 3 + 1 dimensional FLRW Universe with metric
\begin{equation}
g_{\mu \nu} = a^2(\tau) \left[ -d\tau^2 + \left( \delta_{ij} + h_{ij} \right)dx^i dx^j \right],
\end{equation}
where $\tau$ is conformal time and $h_{ij}$ is the transverse and traceless tensor which defines the gravitational waves, and whose equation of motion reads
\begin{equation} \label{gweom}
h_{ij}'' + 2 \frac{a'}{a} h'_{ij} - \Delta h_{ij} = \frac{2}{M_P^2} \Pi_{ij}^{\; \; ab} T_{ab},
\end{equation}
where $M_P  = (8 \pi G)^{-1/2}$ is the reduced Planck mass, $\Pi_{ij}^{\; \; lm} = \Pi_i^l\Pi_j^m - \frac{1}{2}\Pi_{ij}\Pi^{lm}$ is the transverse, traceless projector, $\Pi_{ij} = \delta_{ij} - \partial_i \partial_j / \Delta$, and a prime denotes derivatives with respect to conformal time $\tau$ .   

As we have discussed in the Introduction, our goal is to consider a scenario where the mass of a scalar field $\chi$ goes from a nonvanishing to a vanishing value during inflation. We do so by considering an additional field $\sigma$ which controls the mass of $\chi$ and that behaves as an order parameter in a phase transition describing a symmetry restoration. More specifically, we will consider a system where a field  $\sigma$ and the inflaton $\varphi$ are subject to a potential of the form
\begin{equation}\label{lagphisigma}
\mathcal{L}_{\varphi\sigma} = - \frac{1}{2}\, \partial_\mu \varphi \partial^\mu \varphi - \frac{1}{2}\, \partial_\mu \sigma \partial^\mu \sigma - \frac{\mu}{2}\varphi \, \sigma^2 - \frac{\lambda}{4}\, \sigma^4-V(\varphi)\,,
\end{equation}
where $V(\varphi)$ is some flat potential able to support inflation, $\lambda$ is a dimensionless coupling constant, and $\mu$ is a mass dimension-1 coupling constant. The coupling between $\varphi$ and $\sigma$ would generally take the form $\frac{\mu}{2}\left(\varphi-\varphi_*\right)\,\sigma^2$, where $\varphi_*$ is some constant value crossed by the expectation value of $\varphi$ during inflation. However, we can always set $\varphi_*=0$ by an appropriate shift of $\varphi$. Reference~\cite{Langlois:2004px} also studied a system where a field analogous to $\sigma$ evolved towards a vanishing expectation value during inflation. In that paper, however, that field was not coupled to the inflaton, so that its evolution was more slow than in the present work.

We will assume without loss of generality that $\dot\varphi>0$, so that the term proportional to $\mu$ in the Lagrangian~(\ref{lagphisigma}) behaves like a negative mass squared term for $\sigma$ at early times, triggering symmetry breaking, while at later times it behaves like a positive mass term, enforcing $\sigma=0$. More explicitly, for $\varphi < 0$ the minimum of the potential for $\sigma$ is $\sigma_{min} = \pm \sqrt{-\frac{\mu \varphi}{\lambda}}$, while for $\varphi > 0$, the minimum is $\sigma_{min} = 0$.  We will assume that some earlier inflationary dynamics has chosen one of the two minima, say $\sigma_{min}=+ \sqrt{-\frac{\mu \varphi}{\lambda}}>0$ for the early value of the zero mode of the $\sigma$ field.

Let us now introduce a third field $\chi$, that will be our source of gravitational waves. The field $\chi$ interacts with $\sigma$ through the lagrangian
\begin{equation}
\mathcal{L}_\chi = - \frac{1}{2} \partial_\mu \chi \partial^\mu \chi -
\frac{h^2}{2} \sigma^2 \chi^2,
\end{equation}
where $h$ is a dimensionless coupling constant. If $\sigma$ tracks the minimum of its potential (we will see in Subsection \ref{sec:evolutionzero} under which conditions this requirement is satisfied) then the mass of $\chi$  will be given by
\begin{eqnarray}
  m_\chi = \left\{\def\arraystretch{1.2}%
  \begin{array}{@{}c@{\quad}l@{}}
    \sqrt{-\frac{h^2}{\lambda}\,\mu\,\varphi}  & \text{for} \; t  < t_*  \\
    0 & \text{for} \; t > t_*, \\
  \end{array}\right.
\end{eqnarray}
where $t_*$ corresponds to the time when $\varphi$ crosses $0$. If the inflaton evolves under the usual slow-roll conditions then we can model its time evolution around $t_*$ as
\begin{equation}
\varphi(t) \simeq \dot\varphi_*\, (t - t_*)\, ,
\end{equation}
so that the mass of $\chi$ reads
\begin{equation} \label{chimass}
  m_\chi = \left\{\def\arraystretch{1.2}%
  \begin{array}{@{}c@{\quad}l@{}}
    \Lambda_\chi^\frac{3}{2} \sqrt{t_*-t}  & \text{for} \; t  < t_*  \\
    0 & \text{for} \; t > t_* \\
  \end{array}\right.\,,\qquad\qquad \Lambda_\chi^3 \equiv \frac{h^2\, \mu}{\lambda}\, \dot \varphi_*\,.
\end{equation}

In the following section we will consider how the rolling of $\sigma$ can lead to the production of quanta of $\chi$ which, in their turn, will act as a source of source gravitons. 

\section{Production of gravitational waves}\label{sec:gwprod}%

In order to compute the amplitude of gravitational waves produced by quanta of the $\chi$ field, we must first characterize the production of quanta of the field $\chi$ induced by the time-dependence~(\ref{chimass}) of its mass.  For our subsequent analysis we will also need to study the fluctuations of the field $\sigma$ which follows a similar calculation to that of $\chi$. The fluctuations of $\sigma$ can be studied using the standard formalism of Bogolyubov coefficients. Since the field $\chi$ stay massless after particle production, on the other hand, its superhorizon modes will not be evolving adiabatically after production of its quanta, and we will need a more subtle analysis, which we will present in subsection \ref{subsec:prodchi}.

\subsection{Production of quanta of $\sigma$}\label{subsec:prodsigma}%

\begin{figure}[tbp]
\centering
\includegraphics[width=0.4\textwidth]{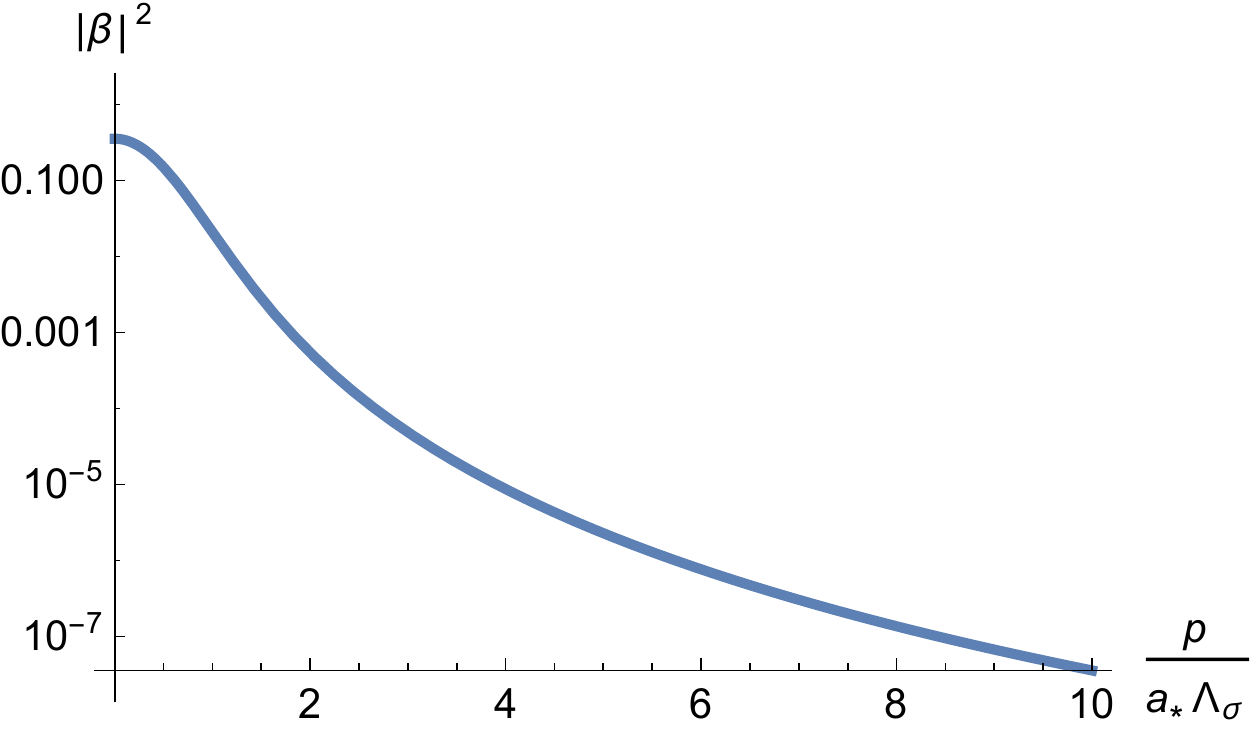}
\caption{The occupation number of $\sigma$ particles $\left|\beta_\sigma\right|^2$ as a function of the momentum $p$ expressed in units of $a_*\,\Lambda_\sigma$.}
\label{fig:betasigma2}
\end{figure}

We will assume that the parameters of the model are such that the field $\sigma$ is heavy (in units of the Hubble scale) for most of the evolution of the system. However, when $\varphi$ crosses zero the field $\sigma$ becomes temporarily massless, and quanta of $\sigma$ are created by resonant effects. We decompose $\sigma$ into a homogeneous $\sigma_0(\tau)$ and perturbed $\delta \sigma(\textbf{x},\tau)$ part
\begin{equation}
\sigma(\textbf{x},\tau) = \sigma_0(\tau) + \delta \sigma(\textbf{x},\tau) \, ,
\end{equation}
and we further decompose the fluctuations as
\begin{equation}
\delta \hat \sigma(\textbf{x},\,\tau) =\frac{1}{a(\tau)} \int \frac{d^3 \textbf{p}}{(2 \pi)^{3/2}} e^{i \textbf{p} \cdot \textbf{x}} \left[ \delta \sigma_\textbf{p}(\tau) \,\hat c_\textbf{p} + \delta \sigma^*_\textbf{-p}(\tau)\, \hat c^\dagger_\textbf{-p} \right]\,,
\end{equation}
where $\hat{c}^{(\dagger)}$ are the ladder operators for  $\delta\hat\sigma$, and where the equation of motion for the canonically normalized fluctuations reads
\begin{equation}
\delta\sigma_\textbf{p}'' + \left[ p^2 - \frac{a''}{a} + m_\sigma^2(\tau) a^2 \right] \delta\sigma_\textbf{p} = 0\,,
\end{equation}
with $m^2_\sigma(\tau < \tau_*) = - 2\, \mu\, \varphi(\tau) $ which shows that as $\varphi$ approaches $0$ the field $\sigma$ becomes massless and the WKB approximation is not a good one for the evolution of its mode functions.  This implies a resonant amplification of the quantum fluctuations of $\sigma$, that we study as it is usual \cite{Kofman:1997yn} by switching to physical time, approximating $\varphi(t) \simeq \dot \varphi_* (t-t_*)$, and neglecting the expansion of the Universe during the period of nonadiabaticity. In this regime, the equation for the mode functions of the rescaled field $\delta\sigma_\textbf{p} = a^{-\frac{1}{2}} \delta \sigma_c$ reads
\begin{equation}\label{eqsigma}
\d\ddot\sigma_c + \left[ \frac{p^2}{a_*^2} -\Lambda_\sigma^3(t-t_*) \right] \delta\sigma_c  = 0\,,
\end{equation}
where we have defined 
\begin{equation}
\Lambda^3_\sigma \equiv 2 \, \mu \,\dot \varphi_*=\frac{2\,\lambda}{h^2}\,\Lambda_\chi^3  .
\end{equation}
The assumption that the expansion of the Universe is negligible during the nonadiabatic period is equivalent to 
\begin{equation}
\Lambda_\sigma\gg H \,,
\end{equation}
which also implies that, as stated above, the mass of $\sigma$ is much larger than the Hubble parameter for most of the time. The solution of eq.~(\ref{eqsigma}) reads
\begin{equation}
\delta\sigma_c (t < t_*) = \sqrt{\frac{\pi\,z}{6\,\Lambda_\sigma}}\, \mathrm{H}^{(1)}_{\frac{1}{3}}\left( \frac{2}{3} z^{\frac{3}{2}}  \right)\,, \hspace{0.5cm} z \equiv\frac{p^2}{a_*^2 \Lambda_\sigma^2} - \Lambda_\sigma (t-t_*)\, ,
\end{equation}
where $H^{(1)}_\nu(z)$ denotes the Hankel function of the first kind and where we have determined the integration constants assuming that the modes of $\sigma$ are in their adiabatic vacuum at early times.  For $t>t_*$ the quanta of $\sigma$ become massive again, and their equation of motion  reads
\begin{equation}
{ \delta\sigma_c'' + \left[ p^2 - \frac{a''}{a} +\mu \, a^2 \varphi \right]\delta\sigma_c = 0\,. }
\end{equation}
Proceeding as we did for $t<t_*$, we obtain
\begin{equation}
\delta\sigma_c(t>t_0) = c_1\,\sqrt{\tilde{z}}\,\mathrm{H}^{(1)}_{\frac{1}{3}}\left( \frac{2}{3}\, \tilde{z}^{\frac{3}{2}}  \right) + c_2\, \sqrt{\tilde{z}} \,\mathrm{H}^{(2)}_{\frac{1}{3}}\left( \frac{2}{3}\, \tilde{z}^{\frac{3}{2}}  \right), \hspace{1.0cm} \tilde{z}\equiv 2^{2/3}\,\frac{p^2}{a_*^2 \Lambda_\sigma^2} +2^{-\frac{1}{3}}\, \Lambda_\sigma \, (t - t_*),
\end{equation}
where $c_1$ and $c_2$ are determined by matching the solution at $t = t_*$,
\begin{align}
&c_1 = i\frac{\pi}{3\times 2^{1/3}}\sqrt{\frac{\pi}{6\,\Lambda_\sigma}} \frac{p^3}{a_*^3\,\Lambda^3_\sigma}\left[\mathrm{H}_{-\frac{2}{3}}^{(1)}\left(\frac{2}{3}\frac{p^3}{a_*^3\,\Lambda^3_\sigma}\right)\,\mathrm{H}_{\frac{1}{3}}^{(2)}\left(\frac{4}{3}\frac{p^3}{a_*^3\,\Lambda^3_\sigma}\right)+\mathrm{H}_{\frac{1}{3}}^{(1)}\left(\frac{2}{3}\frac{p^3}{a_*^3\,\Lambda^3_\sigma}\right)\,\mathrm{H}_{-\frac{2}{3}}^{(2)}\left(\frac{4}{3}\frac{p^3}{a_*^3\,\Lambda^3_\sigma}\right)\right]\nonumber\\
&c_2 = -i\frac{\pi}{3\times 2^{1/3}}\sqrt{\frac{\pi}{6\,\Lambda_\sigma}} \frac{p^3}{a_*^3\,\Lambda^3_\sigma}\left[\mathrm{H}_{\frac{1}{3}}^{(1)}\left(\frac{4}{3}\frac{p^3}{a_*^3\,\Lambda^3_\sigma}\right)\,\mathrm{H}_{-\frac{2}{3}}^{(1)}\left(\frac{2}{3}\frac{p^3}{a_*^3\,\Lambda^3_\sigma}\right)+\mathrm{H}_{\frac{1}{3}}^{(1)}\left(\frac{2}{3}\frac{p^3}{a_*^3\,\Lambda^3_\sigma}\right)\,\mathrm{H}_{-\frac{2}{3}}^{(1)}\left(\frac{4}{3}\frac{p^3}{a_*^3\,\Lambda^3_\sigma}\right)\right]\,.
\end{align}
By matching the above exact solution to the WKB solution for large $\tilde{z}$ we read off the Bogolyubov coefficients as
\begin{align}\label{bogossigma}
&\alpha_\sigma = -i\frac{\pi}{3 \sqrt{2}} \,e^{i (\frac{p^3}{a_*^3 \Lambda_\sigma^3} + \frac{5}{12}\,\pi)}\frac{p^3}{a_*^3\,\Lambda^3_\sigma}\left[\mathrm{H}_{\frac{1}{3}}^{(1)}\left(\frac{4 }{3}\frac{p^3}{a_*^3\,\Lambda^3_\sigma}\right) \,\mathrm{H}_{-\frac{2}{3}}^{(1)}\left(\frac{2}{3}\frac{p^3}{a_*^3\,\Lambda^3_\sigma}\right)+\mathrm{H}_{\frac{1}{3}}^{(1)}\left(\frac{2}{3}\frac{p^3}{a_*^3\,\Lambda^3_\sigma}\right) \,\mathrm{H}_{-\frac{2}{3}}^{(1)}\left(\frac{4}{3}\frac{p^3}{a_*^3\,\Lambda^3_\sigma}\right)\right]\,,\nonumber\\
&\beta_\sigma = i\frac{\pi}{3 \sqrt{2}}\, e^{i (\frac{p^3}{a_*^3 \Lambda_\sigma^3} - \frac{5}{12}\pi)}\frac{p^3}{a_*^3\,\Lambda^3_\sigma} \left[\mathrm{H}_{-\frac{2}{3}}^{(1)}\left(\frac{2}{3}\frac{p^3}{a_*^3\,\Lambda^3_\sigma}\right)\,\mathrm{H}_{\frac{1}{3}}^{(2)}\left(\frac{4}{3}\frac{p^3}{a_*^3\,\Lambda^3_\sigma}\right)+\mathrm{H}_{\frac{1}{3}}^{(1)}\left(\frac{2}{3}\frac{p^3}{a_*^3\,\Lambda^3_\sigma}\right)\,\mathrm{H}_{-\frac{2}{3}}^{(2)}\left(\frac{4}{3}\frac{p^3}{a_*^3\,\Lambda^3_\sigma}\right)\right]\, .
\end{align}
The occupation number $\left|\beta_\sigma\right|^2$ is plotted in figure~\ref{fig:betasigma2}. We will use the above coefficients to determine the effects of the produced $\sigma$ particles in Subsections~\ref{sec:energysigma} and~\ref{sec:sigmaonphi} below. 

\subsection{Production of quanta of $\chi$}\label{subsec:prodchi}%

The analysis of the production of quanta of $\chi$ is similar to that of the previous Subsection, with the additional complication that the $\chi$ particles will remain massless (and the superhorizon modes will therefore not be evolving adiabatically) after the event of particle production. We decompose the field $\chi$  as
\begin{equation}
\hat \chi(\textbf{x},\,\tau) =\frac{1}{a(\tau)} \int \frac{d^3 \textbf{p}}{(2 \pi)^{3/2}} e^{i \textbf{p} \cdot \textbf{x}} \left[ \chi_\textbf{p}(\tau) \,\hat a_\textbf{p} + \chi^*_\textbf{-p}(\tau)\, \hat a^\dagger_\textbf{-p} \right]\,,
\end{equation}
where the (canonically normalized) mode functions satisfy
\begin{equation}
\chi_\textbf{p}'' + \left[ p^2 - \frac{a''}{a} + m^2_\chi(\tau)\, a^2 \right] \chi_\textbf{p} = 0\,,
\end{equation}
with $m_\chi(\tau)$ given by eq.~(\ref{chimass}). We will assume that the parameters of the system are such that $m_\chi$ evolves adiabatically, $|(a\,{m}_\chi)'|\ll a^2\,m_\chi^2$ for most of the time, and that the period in which the adiabaticity condition is violated, close to the time when $m_\chi=0$, is much shorter than an Hubble time, which implies the condition
\begin{align}
\Lambda_\chi\gg H\,.
\end{align}

We can now determine the mode functions during the nonadiabatic regime by switching from conformal to physical time, neglecting the expansion of the Universe during this epoch, and introducing the rescaled field $\chi = a^{-\frac{1}{2}} {\chi}_c$. Then the equation for $\chi_c$ simplifies to
\begin{equation}\label{eqchi}
\ddot{{\chi}}_c + \left[ \frac{p^2}{a_*^2} +\Lambda_\chi^3 \,(t_* - t) \right]\,{\chi}_c = 0\,,
\end{equation}
whose solution for $t < t_*$, reducing to the adiabatic vacuum at early times, can be written as
\begin{equation}\label{chisol_early}
\chi_c(t < t_*) = \sqrt{\frac{\pi\, z}{6\,\Lambda_\chi}} \mathrm{H}^{(1)}_{\frac{1}{3}}\left( \frac{2}{3} z^{\frac{3}{2}}  \right)\,, \hspace{1.0cm} z \equiv \frac{p^2}{a_*^2\,\Lambda_\chi^2} + \Lambda_\chi\left(t_*-t\right)\,.
\end{equation}
For $\tau>\tau_*$ the scalar is massless, so that its mode functions are given by
\begin{equation} \label{eomml}
\chi_c(\tau > \tau_*) = c_+\,\frac{e^{-i p \tau}}{\sqrt{2\,p}}\left( 1 - \frac{i}{p \,\tau} \right) + c_-\,\frac{e^{i p \tau}}{\sqrt{2\,p}}\left( 1 + \frac{i}{p \,\tau} \right)\,,
\end{equation}
where the constants $c_+$ and $c_-$ are determined by imposing continuity of $\chi_c$ and of its first derivative at $\tau_*$, so that
\begin{align}\label{cp1}
\hspace{-1.0cm}
c_+ = \sqrt{\frac{\pi a_*^2\,H^4}{12\, p\, \Lambda_\chi^3}}\, e^{-\frac{2\,p^3}{3\,a_*^3\,H^3}i + \frac{5\pi}{12}i} \,&\left[\left(-1-i\,\frac{p}{a_*\,H}+\left(\frac{p}{a_*H}\right)^2\right)\, H_{\frac{1}{3}}^{(1)}\left(\frac{2\,p^3}{3\,a_*^3\,\Lambda_\chi^3}\right)\right.\nonumber\\
&\qquad\qquad-\left.\left(\frac{p}{a_*H} +i \left(\frac{p}{a_*H}\right)^2 \right)\, H_{-\frac{2}{3}}^{(1)}\left(\frac{2\,p^3}{3\,a_*^3\Lambda_\chi^3}\right)\right]\,,\nonumber\\
c_- = \sqrt{\frac{\pi a_*^2\,H^4}{12\, p\, \Lambda_\chi^3}}\, e^{ i \frac{p}{a_*H} -\frac{2\,p^3}{3\,a_*^3\,H^3}i + \frac{5\pi}{12}i} &\left[\left(-1+\frac{p}{a_*H}+i\,\left(\frac{p}{a_*H}\right)^2\right)\, H_{\frac{1}{3}}^{(1)}\left(\frac{2\,p^3}{3\,a_*^3\Lambda_\chi^3}\right)\right.\nonumber\\
&\left.\qquad\qquad + \left(i\,\frac{p}{a_*H} -\left(\frac{p}{a_*H}\right)^2 \right) \, H_{-\frac{2}{3}}^{(1)}\left(\frac{2\,p^3}{3\,a_*^3\Lambda_\chi^3}\right)\right]\,.
\end{align}
Note that in the Minkowski limit $H\to 0$, $\tau_*=-H^{-1}\to\infty$ the coefficients $c_+$ and $c_-$ converge to the Bogolyubov coefficients, respectively $\alpha$ and $\beta$, in Minkowski space provided in Appendix \ref{app:min}.  In a de Sitter background, however, the interpretation $c_+$ and $c_-$ as Bogolyubov coefficients associated to the occupation number of $\chi$ particles is not  rigorous, since for $\tau>\tau_*$ the superhorizon modes of $\chi$ are not evolving adiabatically. Moreover, we do not expect the coefficients~(\ref{cp1}) to be accurate for modes that were superhorizon, $k/a_*<H$, at the time $\tau_*$, since the solution~(\ref{chisol_early}) has been found under the assumption that the Hubble parameter is negligibly small, so that those modes are not accounted for.  Since $\Lambda_\chi\gg H$, however, we believe that the error is negligible, and, as we will show below, we will still be able to  effectively study the production of $\chi$ through a subtraction method.

\subsection{Production of gravitational waves}\label{subsec:gws}

We now calculate the spectrum of gravitational waves sourced by the field $\chi$.  The graviton $h_{ij}$ obeys eq.~(\ref{gweom}), where the spatial components of the stress-energy tensor for the $\hat\chi$ operator, expanding to first order in $h_{ab}$, is given by
\begin{equation}
T_{ab}=\partial_a\hat\chi\,\partial_b\hat\chi+\hat{h}_{ab}\left[\frac{1}{2}{\hat\chi}'{}^2-\frac{1}{2}\left(\nabla\hat\chi\right)^2-a^2\,V(\hat\chi)\right]+\dots\,,
\end{equation}
where the dots denote terms that are second or higher order in $\hat{h}_{ij}$ and the terms that are proportional to $\delta_{ij}$ and are projected out by $\Pi_{ij}^{\; \; ab}$.

Next, we write $\hat{h}_{ij}$ as $\hat{h}_{ij}^{(0)}+\hat{h}_{ij}^{(1)}$, where $\hat{h}_{ij}^{(0)}$ is a solution of the homogeneous equation $\hat{h}_{ij}'' + 2 \frac{a'}{a} \,\hat{h}'_{ij} - \Delta \hat{h}_{ij} =0$. Therefore, to leading order, the equation for $\hat{h}_{ij}^{(1)}$ is solved by
\begin{equation}\label{solds}
\hat{h}_{ij}^{(1)}({\bf p},\,t)= \frac{2}{M_P^2} \int d\tau' G_p(\tau,\,\tau') \,\Pi_{ij}^{\; \; ab}(\textbf{p}) \left(\partial_a\hat\chi\,\partial_b\hat\chi+\hat{h}^{(0)}_{ab}\,\left[\frac{1}{2}\hat\chi'{}^2-\frac{1}{2}\left(\nabla\hat\chi\right)^2-a^2\,V(\hat\chi)\right]\right)\left({\bf p},\,\tau'\right)\, ,
\end{equation}
where the Green's function for the graviton reads
\begin{equation}
G_k(\tau,\tau') = \frac{1}{k^3  {\tau'}^2} \left[ (1 + k^2 \tau \tau') \sin (k (\tau - \tau')) + k(\tau' - \tau) \cos (k(\tau-\tau')) \right] \Theta(\tau - \tau').
\end{equation}

Using the fact that $h_{ij}^{(0)}$ and $\chi$ are uncorrelated, we find the graviton correlator  as the sum of three terms
\begin{equation}
\braket{\left(h_{ij}^{(0)}(\textbf{k},\,\tau)+h_{ij}^{(1)}(\textbf{k}',\,\tau)\right)\,\left(h_{ij}^{(0)}(\textbf{k},\,\tau)+h_{ij}^{(1)}(\textbf{k}',\,\tau)\,\right)}=\frac{2\,\pi^2}{k^3}\delta(\textbf{k}+\textbf{k}')\,\left({\cal P}_T^{00}+{\cal P}_T^{11}+2\,{\mathrm {Re}}\left\{{\cal P}_T^{01}\right\}\right)\,,
\end{equation}
where $\frac{2\,\pi^2}{k^3}\delta(\textbf{k}+\textbf{k}')\,{\cal P}_T^{00}=\braket{h_{ij}^{(0)}(\textbf{k},\,\tau)  h_{ij}^{(0)}(\textbf{k}',\,\tau)}=\frac{2\,\pi^2}{k^3}\delta(\textbf{k}+\textbf{k}')\,\frac{2}{\pi^2}\,\frac{H^2}{M_P^2}$ is the standard vacuum contribution, while
\begin{align} \label{P11}
&\frac{2\,\pi^2}{k^3}\,\delta(\textbf{k}+\textbf{k}')\,{\cal P}_T^{11} =  \frac{1}{2\, \pi^3\, M_P^4} \int d\tau'\, G_k(\tau,\,\tau') \int d\tau''\, G_{k'}(\tau,\,\tau'') \,\Pi_{ij}^{\; \; ab}(\textbf{k})\, \Pi_{ij}^{\; \; cd}(\textbf{k}') \nonumber \\ 
&\times \int d^3\textbf{p}\,d^3\textbf{p}'\, p_a\,(\textbf{k}_b - \textbf{p}_b)\, p'_c\,(\textbf{k}_d' - \textbf{p}_d')\, \braket{\hat \chi(\textbf{p},\,\tau')\,\hat \chi(\textbf{k}-\textbf{p},\tau')\,\hat \chi(\textbf{p}',\,\tau'')\,\hat \chi(\textbf{k}'-\textbf{p}',\tau'')}\,,
\end{align}
is the contribution that originates from the term proportional to $\langle  \chi^4\rangle$ in $\langle h_{ij}^{(1)}(\textbf{k},\,\tau)\,h_{ij}^{(1)}(\textbf{k}',\,\tau)\rangle$.

Finally,
\begin{align} \label{P01}
&\frac{2\,\pi^2}{k^3}\,\delta(\textbf{k}+\textbf{k}')\,{\cal P}_T^{01} =  \frac{1}{8\, \pi^3\, M_P^2} \int d\tau'\, G_{k'}(\tau,\,\tau') \,\Pi_{ij}^{\; \; ab}(\textbf{k}')\,  \int d^3\textbf{p}\,d^3\textbf{p}'\nonumber \\ 
&\times\left\langle\hat h^{(0)}_{ij}({\bf k},\,\tau)\,\hat{h}^{(0)}_{ab}({\bf k}'-{\bf p}-{\bf p}',\,\tau') \left[\hat\chi'(\textbf{p},\,\tau')\,\hat\chi'(\textbf{p}',\,\tau')+\left({\bf p}\cdot{\bf p}'-m_\chi^2\,a(\tau')^2\right)\,\hat\chi(\textbf{p},\,t')\,\hat\chi(\textbf{p}',\,\tau')\right]\right\rangle\,,
\end{align}
comes from the cross term between $h_{ij}^{(0)}$ and the part proportional to $h_{ij}^{(0)}$ in eq.~(\ref{solds}). Notice that, since we are evaluating the amplitude of the tensors produced at times $t>t_*$, we will set $m_\chi=0$.

The existence of the two contributions ${\cal P}_T^{11}$ and ${\cal P}_T^{01}$ can also be derived in the context of the in-in formalism, and originates from the two different diagrams presented in~\cite{Carney:2012pk}.

\subsubsection{Obtaining finite quantities} \label{sec:finite}

To evaluate ${\cal P}_T^{01}$ and ${\cal P}_T^{11}$ we use Wick's theorem, so that we have to evaluate correlators of the form $\langle\hat\chi(\bk_1,\,\tau_1)\,\hat\chi(\bk_2,\,\tau_2)\rangle$, that need to be renormalized. If we were to perform this calculation on a Minkowski background, we would have a straightforward and physically transparent way of performing such a renormalization. On a Minkowski background, in fact, the frequency of the field $\hat\chi$ (which is just its momentum) evolves adiabatically after $\tau_*$, so that we can fully use the formalism of Bogolyubov coefficients. This means that we would decompose the field in terms of new creation/annihilation operators $\hat{b}_\bk^{(\dagger)}$, where the $\hat{b}_\bk$ operator multiplies the positive frequency component of the mode functions of $\hat\chi$ for times $\tau>\tau_*$, when the frequency of the modes of $\hat\chi$ is adiabatically evolving. Then  we would normal order the $\hat\chi(\bk_1,\,\tau_1)\,\hat\chi(\bk_2,\,\tau_2)$ operator in terms of the $\hat{b}_\bk^{(\dagger)}$ ladder operators, and not in terms of the original $\hat{a}_\bk^{(\dagger)}$ ones, that were used to quantize $\hat\chi$ for $\tau<\tau_*$. This means that observers born after $\tau=\tau_*$ would renormalize away the vacuum fluctuations of the mode functions defined for $\tau>\tau_*$ by normal ordering the operators $\hat{b}_\bk^{(\dagger)}$, that correspond to their notion of a particle.  Writing the relationship between the $b_\bk^{(\dagger)}$ and the $a_\bk^{(\dagger)}$ operators as
\begin{equation}\label{decohat}
\hat{b}_\bk(\tau)=\alpha(\bk,\,\tau)\,\hat{a}_\bk+\beta^*(-\bk,\,\tau)\,\hat{a}_{-\bk}^\dagger\,,
\end{equation}
we would obtain
\begin{align}\label{twoptchi1}
\langle\hat\chi(\bp,\,\tau')\hat\chi(\bq,\,\tau'')\rangle&=\frac{\delta^{(3)}(\bp+\bq)}{2\,\sqrt{\omega_\bp(\tau')\,\omega_\bp(\tau'')}} \left[\left(e^{i\int_{\tau'}^{\tau''}\omega_\bp } \,\beta^*(-\bp,\,\tau')\beta(-\bp,\,\tau'')+ {\rm h.c.}\right)\right.\\
& +\left.\left( e^{-i \int^{\tau'} \omega_\bp -i \int^{\tau''} \omega_\bp } \alpha({\bf p}, \tau') \beta^*({\bf p}, \tau'') +\left (\tau'\leftrightarrow \tau'', {\mathrm {h.c.}}\right) \right)  \right]\,.\nonumber
\end{align}

As we will also discuss in Appendix \ref{app:min},  the prescription~(\ref{twoptchi1}) above is equivalent to setting
\begin{equation}\label{corrsubtr1}
\braket{\chi(\textbf{p},\,\tau')\chi(\textbf{q},\,\tau'')} = \delta^{(3)}(\textbf{p} + \textbf{q})\, \left[ \chi({\bf p},\,\tau')\,\chi({\bf p},\,\tau'')- \tilde\chi({\bf p},\,\tau')\,\tilde\chi({\bf p},\,\tau'')\right],
\end{equation}
where, in the case of particles on a Minkowskian background, $\tilde{\chi}({\bf p},\,\tau)=\frac{e^{-i\,p\,\tau}}{\sqrt{2\,p}}$ corresponds to the mode functions in the absence of particle creation, $\Lambda_\chi\to 0$. In its turn, this means that the procedure presented above corresponds precisely to that of adiabatic regularization, where one subtracts from the UV-divergent propagator its adiabatic part to obtain a finite result.

Let us now go back to the production of quanta of $\hat\chi$ in de Sitter space. As discussed above, since we are talking about massless particles in de Sitter space, we have $\omega^2=k^2-2/\tau^2$ that for $k\lesssim -1/\tau$ is not evolving adiabatically. Therefore the prescription~(\ref{twoptchi1}) cannot be applied to this case. However, the prescription~(\ref{corrsubtr1}), that on a Minkowski background is equivalent to (\ref{twoptchi1}), {\em can} be applied to our de Sitter background, once we set
\begin{align}
\tilde\chi({\bf k},\,\tau)=\chi({\bf k},\,\tau)\Big|_{\Lambda_\chi\to 0}.
\end{align}

Based on the above considerations, we will use
\begin{equation}\label{corrsubtrds}
\braket{\chi(\textbf{p},\,\tau')\chi(\textbf{q},\,\tau'')} = \delta^{(3)}(\textbf{p} + \textbf{q})\, \left[ \chi({\bf p},\,\tau')\,\chi({\bf p},\,\tau'')- \tilde\chi({\bf p},\,\tau')\,\tilde\chi({\bf p},\,\tau'')\right]\,,
\end{equation}
where the function $\chi(\bp,\,\tau)$ is given by eq.~(\ref{eomml}) with the integration constants given by eq.~(\ref{cp1}), whereas $\tilde\chi({\bf p},\,\tau)$ is obtained by setting $\Lambda\to 0$ in  $\chi(\bp,\,\tau)$, so that
\begin{equation} \label{adiabaticchi}
\tilde{\chi}(\textbf{p},\,\tau) = b_+ \frac{e^{-ip\tau}}{\sqrt{2\,p}}\left[ 1 - \frac{i}{p\,\tau} \right] + b_- \frac{e^{ip\tau}}{\sqrt{2\,p}}\left[ 1 + \frac{i}{p\,\tau} \right]\,,
\end{equation}
with
\begin{equation} \label{subcp}
b_+ = 1 -\frac{i}{y} -\frac{a_*^2\,H^2}{2\, p^2}\,,\qquad b_- = -\frac{a_*^2\,H^2}{2\, p^2}\,e^{\frac{2p}{a_*H}i}.
\end{equation}
This prescription is analogous to that used for instance in~\cite{Barnaby:2009mc}.

\subsubsection{Evaluation of the tensor power spectrum}

Let us now evaluate ${\cal P}_T^{11}$ and ${\cal P}_T^{01}$.

\begin{itemize}

\item ${\cal P}_T^{11}$. To  calculate ${\cal P}_T^{11}$ we need the following expression, that allows us to compute the factor proportional to the transverse-traceless projectors
\begin{equation} \label{proj}
\Pi_{ij}^{\; \; ab}(\textbf{k})\, \Pi_{ij}^{\; \; cd}(\textbf{k}') \,p_a(\,\textbf{k}_b - \textbf{p}_b)\, (\textbf{k}_c - \textbf{p}_c)\, p_d = \frac{1}{2} \left( p^2 - \frac{(\textbf{p}\cdot\textbf{k})^2}{k^2} \right)^2\,.
\end{equation}

After taking the limit $\tau\to 0$, so that we evaluate the effects at the end inflation, when the relevant scales are well outside of the horizon, we obtain
\begin{align}\label{P11int}
& {\cal P}_T^{11} =  \frac{H^4 }{2\,  \pi^3 \,k^3\, M_P^4}  \int  d^3\textbf{p} \; \left( p^2 - \frac{(\textbf{p}\cdot\textbf{k})^2}{k^2} \right)^2\nonumber\\
&\times  \int d\tau' d\tau'' \left[ - \sin(k \tau') + k \tau' \cos(k \tau') \right] \left[ - \sin(k \tau'') + k \tau'' \cos(k \tau'') \right] \nonumber \\ 
&\times  \left[ \chi_\textbf{p}(\tau') \chi^*_\textbf{p}(\tau'') - \tilde{\chi}_\textbf{p}(\tau') \tilde{\chi}^*_\textbf{p}(\tau'')  \right] \left[ \chi_{\textbf{k}-\textbf{p}}(\tau') \chi^*_{\textbf{k}-\textbf{p}}(\tau'') - \tilde{\chi}_{\textbf{k}-\textbf{p}}(\tau') \tilde{\chi}^*_{\textbf{k}-\textbf{p}}(\tau'')  \right].
\end{align}

We have integrated the above expression numerically and a plot for $\Lambda_\chi = 10\,H,\,20\,H$ and $30\,H$ is shown in the left panel of Figure \ref{fig:l10} as a function of $-k\,\tau_*$.   In the right panel of Figure \ref{fig:l10} we plot the amplitude of ${\cal P}_T^{11}$  at the peak $-k\,\tau_* \simeq 3$ as a function of $\Lambda_\chi/H$. That figure shows that the amplitude of the spectrum ${\cal P}_T^{11}$ at its peak scales, for $\Lambda_\chi\gg H$,  as 
\begin{equation} \label{ps11value}
P_h(k) = 2.5\times 10^{-6}\, \frac{H^4}{M_P^4} \frac{\Lambda_\chi^5}{H^5}\,.
\end{equation}

\begin{figure}[tbp]
\centering
\includegraphics[width=0.4\textwidth]{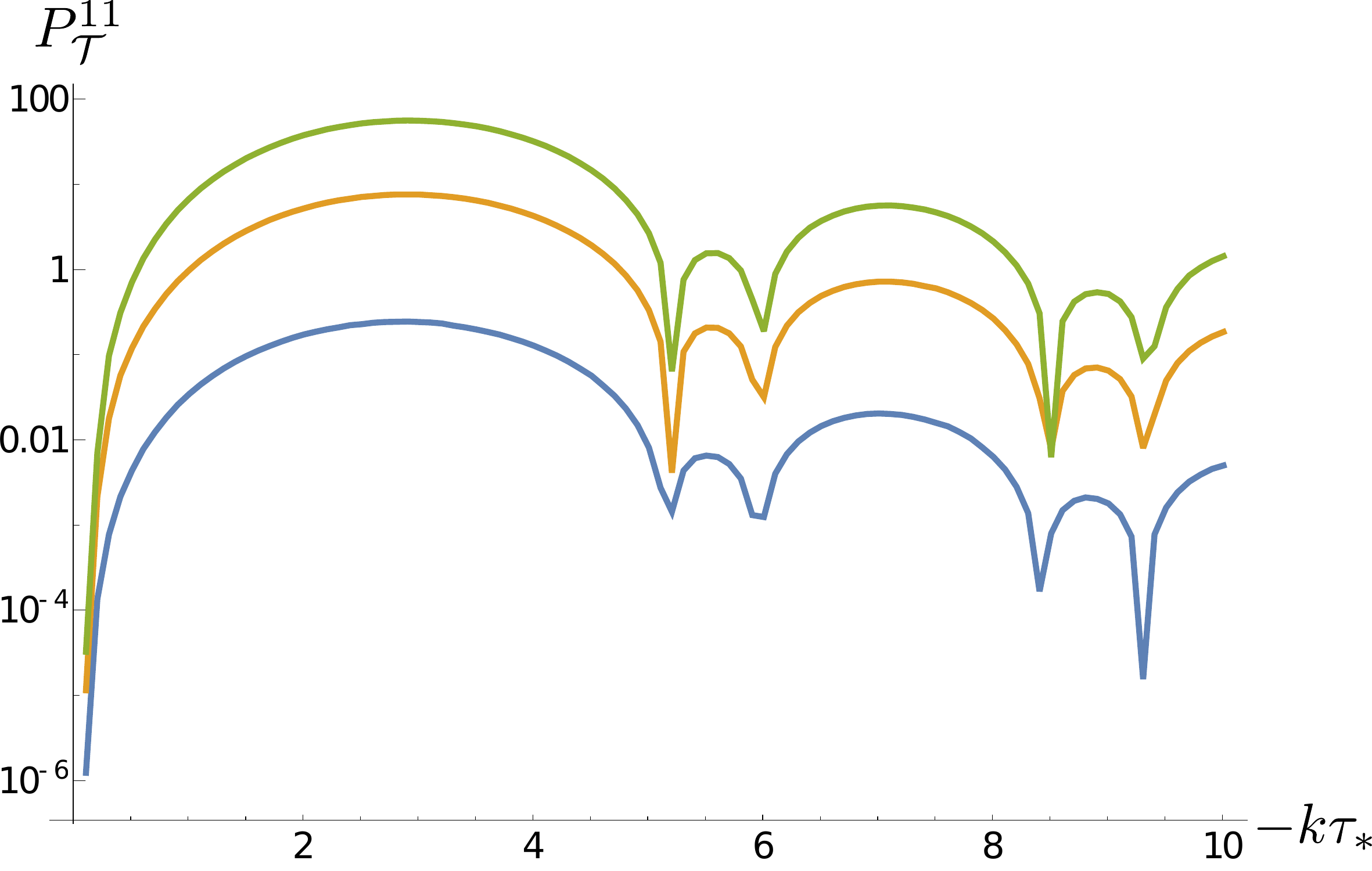}
\includegraphics[width=0.4\textwidth]{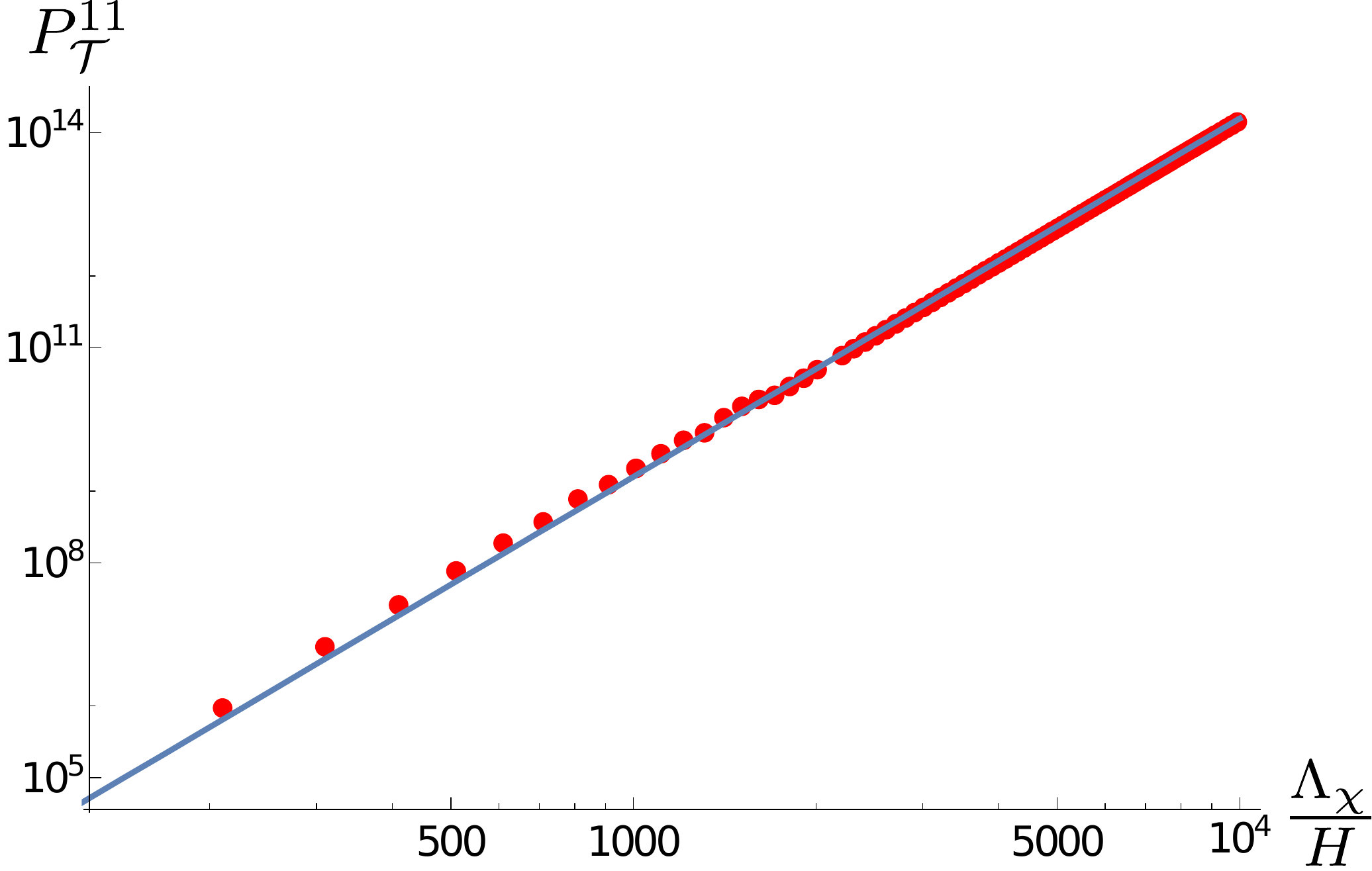}
\caption{Left: Numerical plot of $\frac{M_P^4}{H^4}\,{\cal P}_T^{11}$ as a function of $-k\,\tau_*$ for (top to bottom) $\Lambda_\chi = 30\,H,\,20\,H,\,10\,H$. Right: The amplitude of ${\cal P}_T^{11}$ at its peak, $-k\,\tau_* \simeq 3$, as a function of $\Lambda_\chi/H$. The solid line corresponds to the fit ${\cal P}_T^{11}(-k\,\tau_*=3)= 2.5\times 10^{-6}\, \frac{\Lambda_\chi^5}{M_P^4\,H}$, the red bullets correspond to numerical evaluation of the integral~(\ref{P11int}). }
\label{fig:l10}
\end{figure}

\item ${\cal P}_T^{01}$. The relevant correlator is computed from
\begin{align} \label{P01int}
& \frac{2\,\pi^2}{k^3}\,\delta(\bk_1+\bk_2)\,\times\left(2\,{\mathrm {Re}}\left\{{\cal P}_T^{01}\right\} \right)= -\frac{1}{M_P^2} \int d \tau'  \int \frac{d^3\textbf{q}\, d^3\textbf{q}'}{(2 \pi)^3} \nonumber\\
&\times\big[ G_{k'}(\tau,\tau') \braket{h^{(0)}_{ij}(\textbf{k},\tau)\,h^{(0)}_{ij}(\textbf{k}' - \textbf{q}-\textbf{q}',\tau')} +G_{k}(\tau,\tau')\, \braket{h^{(0)}_{ij}(\textbf{k} - \textbf{q}-\textbf{q}',\tau')\, h^{(0)}_{ij}(\textbf{k}',\tau)} \big] 
\nonumber \\ 
& \times\left[  \chi'(\textbf{q},\tau')\, \chi'(\textbf{q}',\tau') + (\textbf{q} \cdot \textbf{q}') \chi(\textbf{q},\tau')\, \chi(\textbf{q}',\tau') \right]\,,
\end{align}
where the graviton correlator is given by
\begin{equation}
\braket{h^{(0)}_{ij}(\textbf{k},\tau)\,h^{(0)}_{ij}(\textbf{k}',\tau')} =  \frac{4\, \delta^{(3)}(\textbf{k} + \textbf{k}') }{a(\tau)\,a(\tau') \,M_P^2 \,k} \left( 1 + \frac{i(\tau - \tau')}{k\, \tau\, \tau'} + \frac{1}{k^2\, \tau\, \tau'} \right) e^{-i\,k\,(\tau - \tau')}\,.
\end{equation}

\begin{figure}[tbp]
\centering
\includegraphics[width=0.4\textwidth]{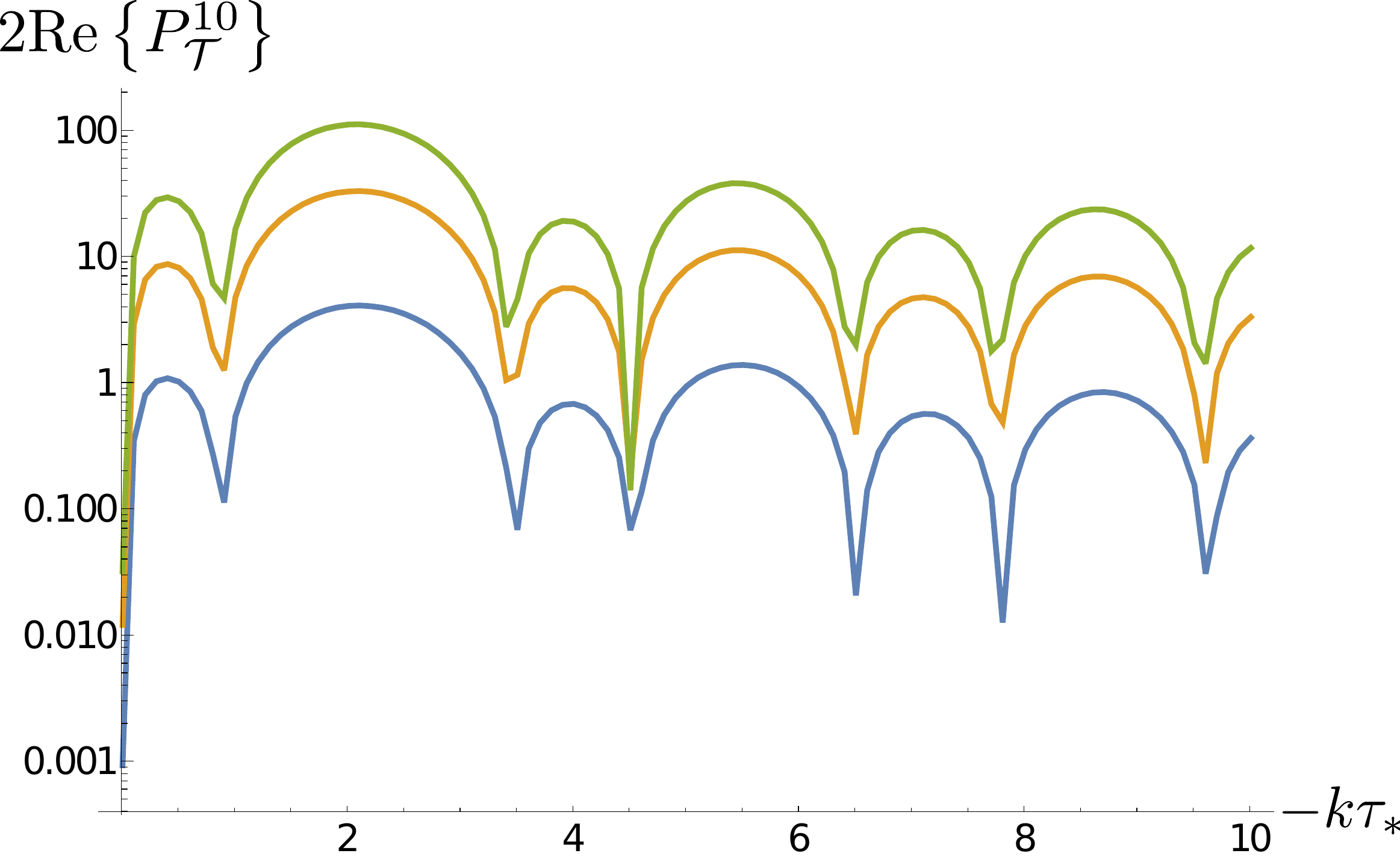}
\hspace{.2cm}
\includegraphics[width=0.4\textwidth]{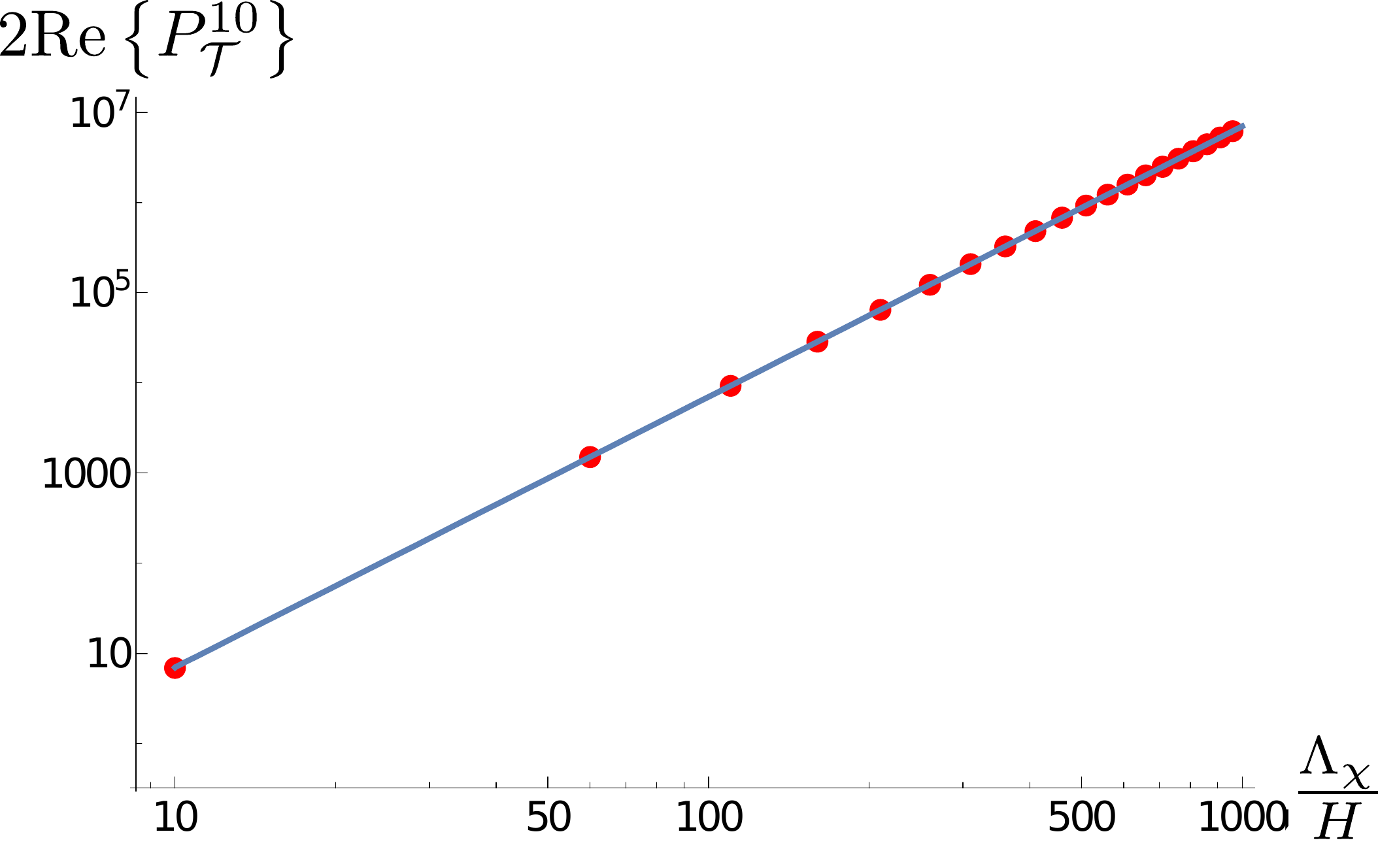}
\caption{Left: Numerical plot of $\frac{M_P^4}{H^4}\,2\,{\mathrm Re}\left\{{\cal P}_T^{01}\right\}$ as a function of $-k\,\tau_*$ for (top to bottom) $\Lambda_\chi = 30\,H,\,20\,H,\,10\,H$. Right: The amplitude of $\frac{M_P^4}{H^4}\,2\,{\mathrm Re}\left\{{\cal P}_T^{01}\right\}$ at its peak, $-k\,\tau_*\simeq 2$, as a function of $\Lambda_\chi/H$. The solid line corresponds to the analytical fit ${\cal P}_T^{11}(-k\,\tau_*=2)= 4\times 10^{-3}\, \frac{\Lambda_\chi^3\,H}{M_P^4}$, the red bullets correspond to numerical evaluation of the integral~(\ref{P01int}). }
\label{fig:l20}
\end{figure}

Taking  $\tau \rightarrow 0$, we obtain an expression for ${\cal P}_T^{01}$ that can be integrated numerically for various values of $\Lambda_\chi/H$. Spectra for $\Lambda_\chi = 10\,H,\,20\,H$ and $30\,H$ are given in the left panel of Figure~\ref{fig:l20}. As we see, the spectra have a peak at $-k\,\tau_*\simeq 2$. In the right panel of Figure~\ref{fig:l20}  we show the amplitude of $2\,{\mathrm {Re}}\left\{{\cal P}_T^{01}\right\}(-k\,\tau_*=2)$ as a function of $\Lambda_\chi$. The numerical fit shows that the amplitude of $2\,{\mathrm {Re}}\left\{{\cal P}_T^{01}\right\}$ at its peak is well approximated by
\begin{equation} \label{ps10value}
2\,{\mathrm {Re}}\left\{{\cal P}_T^{01}\right\}\sim 4 \times 10^{-3} \frac{\Lambda_\chi^3 H}{M_P^4}\,.
\end{equation}
Since this is a factor $\sim H^2/\Lambda_\chi^2$ smaller than the amplitude of ${\cal P}_T^{11}$, we conclude that ${\cal P}_T^{01}$ gives a negligible contribution to the spectrum of gravitational waves produced by the quanta of $\chi$.

\end{itemize}

Before concluding this Section we note that the scalings ${\cal P}_T^{01}\propto \Lambda_\chi^3$ and ${\cal P}_T^{11}\propto \Lambda_\chi^5$ agree with the scalings found on a Minkowski background, see Appendix~\ref{app:min}, where the Bogolyubov coefficients are well defined. This finding lends support to the validity of the subtraction procedure outlined above.

\section{Constraints on the parameter space of the model}
\label{sec:constraints}
%

We have seen in the previous Section that the amplitude of gravitational waves induced by the quanta of $\chi$ goes, for large values of $\Lambda_\chi$, as $\sim 10^{-6}\,\frac{\Lambda_\chi^5}{M_P^4\,H}$, with  $\Lambda_\chi^3 \equiv \frac{ h^2\, \mu}{\lambda}\, \dot \varphi_*$. The energy scale $\Lambda_\chi$ can be in principle very large and might lead to a very large amplitude of induced gravitational waves. In this section we focus on the specific model described in Section \ref{sec:setup} to evaluate the constraints on the parameter space of this scenario and the maximum possible amplitude of ${\cal P}_T^{11}$.

Consistency of our analysis will require a number of conditions that we will now detail.

\subsection{Perturbativity of Coupling Constants}

This condition is simple:
\begin{equation}
h < 1 \, , \hspace{1.0cm} \lambda < 1 \, .
\end{equation}

\subsection{Masses of $\sigma$ and $\chi$}

We require that $\sigma$ and $\chi$ be massive and cosmologically irrelevant for most of the evolution of the system, with the exception of a short period (much less than one efold) around the time $t_*$. Since the masses of those fields are proportional to $\Lambda_{\sigma,\chi}^3\,|t-t_*|$, this condition is equivalent to requiring
\begin{align}
\Lambda_\sigma\gg H\,,\qquad \Lambda_\chi\gg H\,.
\end{align}

\subsection{Evolution of the zero modes} \label{sec:evolutionzero}

We require the validity of the approximate dynamics described in Section~\ref{sec:setup}. Therefore the dynamics of the zero mode $\varphi$ should not be affected significantly by the interactions with $\sigma$, and $\sigma$ should follow the instantaneous minimum of its potential, $\sigma\simeq \sqrt{-\mu\,\varphi/\lambda}$.

The equations of motion for the zero modes $\varphi_0$ and $\sigma_0$ are
\begin{align}
&\ddot\varphi_0+3\,H\,\dot\varphi_0+\frac{\mu}{2}\,\sigma_0^2+V'(\varphi)=0\,,\nonumber\\
&\ddot\sigma_0+3\,H\,\dot\sigma_0+\mu\,\varphi_0\,\sigma_0+\lambda\,\sigma_0^3=0\,,\nonumber\\
&H^2=\frac{1}{3\,M_P^2}\left(\frac{\dot\varphi_0^2}{2}+\frac{\dot\sigma_0^2}{2}+\frac{\mu}{2}\,\varphi_0\,\sigma_0^2+\frac{\lambda}{4}\,\sigma_0^4+V(\varphi)\right)\, .
\end{align}
We parametrize the potential near $\varphi=0$ as 
\begin{align}
V(\varphi) \simeq 3 \bar{H}^2 M_P^2 - 3 \sqrt{2 \epsilon} \bar{H}^2 M_P \, \varphi+3\,\frac{\eta}{2}\bar{H}^2\,\varphi^2 \, ,
\end{align}
where $\epsilon$ and  $\eta$ are the (constant) slow roll parameters. Then, performing the following redefinitions
\begin{align}
& \tilde{H} = \frac{H}{\bar{H}} \, ,\quad \tilde{\phi} = \frac{\varphi_0}{M_P} \, , \quad\tilde{\sigma} = \sigma \sqrt{\frac{\lambda}{\mu M_P}} \, , \nonumber \\
& g_\sigma\equiv\frac{\mu\,M_P}{\bar{H}^2} \, , \quad g_\phi\equiv\frac{\mu^2}{\lambda\,\bar{H}^2} \, ,
\end{align}
we can rewrite the background equations as
\begin{align}\label{eq:rescaled_background}
&{\tilde\phi}''+3\,\tilde{H}\,{\tilde\phi}'+\frac{g_\phi}{2}\,\tilde\sigma^2-3\,\sqrt{2\,\epsilon}+3\,\eta\,\tilde\phi=0\,,\nonumber\\
&\tilde\sigma''+3\,\tilde{H}\,\tilde\sigma'+g_\sigma\,\tilde\phi\,\tilde\sigma+g_\sigma\,\tilde\sigma^3=0\,,\nonumber\\
&\tilde{H}^2=1+\frac{1}{3}\left(\frac{{\tilde\phi}'{}^2}{2}+\frac{g_\phi}{g_\sigma}\,\frac{\tilde\sigma'{}^2}{2}+\frac{g_\phi}{2}\,\tilde\phi\,\tilde\sigma^2+\frac{g_\phi}{4}\,\tilde\sigma^4-\sqrt{2\,\epsilon}\,3\,\tilde\phi+\frac{3}{2}\,\eta\,\tilde\phi^2\right)\,,
\end{align}
where a prime denotes a derivative with respect to $\bar H \,t$.  Then the conditions 
\begin{align} \label{eq:smallquan}
g_\phi\ll 6\,|\eta|\,,\qquad g_\sigma\sqrt{2\epsilon}\gg 1\,,
\end{align}
are sufficient to guarantee that $\tilde\phi\simeq \sqrt{2\,\epsilon}\,\bar{H}\,(t-t_*)$, $\tilde\sigma\simeq \sqrt{-\tilde\phi}$, and $\tilde{H}\simeq 1$ provide good approximations to the actual solutions for our system for a few efoldings around $t=t_*$. The equality $g_\sigma=\Lambda_\sigma^3/(2\,\sqrt{2\,\epsilon}\,\bar{H}^3)$ implies that the condition $g_\sigma\sqrt{2\epsilon}\gg  1$ is identically verified since we are already assuming $\Lambda_\sigma\gg H$.

Therefore we conclude that the only new condition required for the background dynamics is just $g_\phi\ll  6\,|\eta|$, which is equivalent to
\begin{align}\label{eq:gphismall}
\frac{\mu^2}{\lambda\,{H}^2}\ll  6\,|\eta|\,.
\end{align}

\begin{figure}[tbp]
\centering
\includegraphics[width=0.31\textwidth]{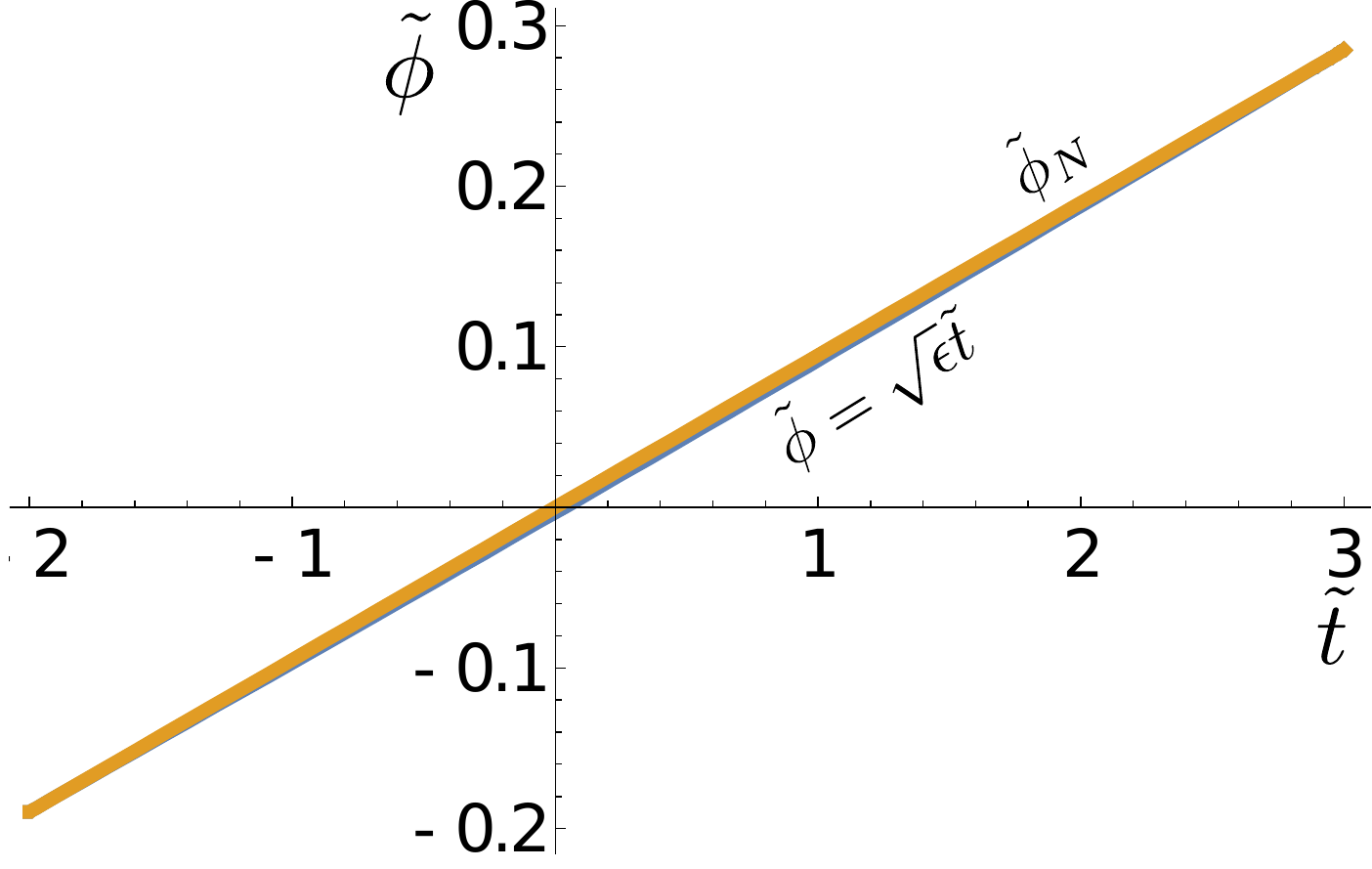}
\hspace{.1cm}
\includegraphics[width=0.31\textwidth]{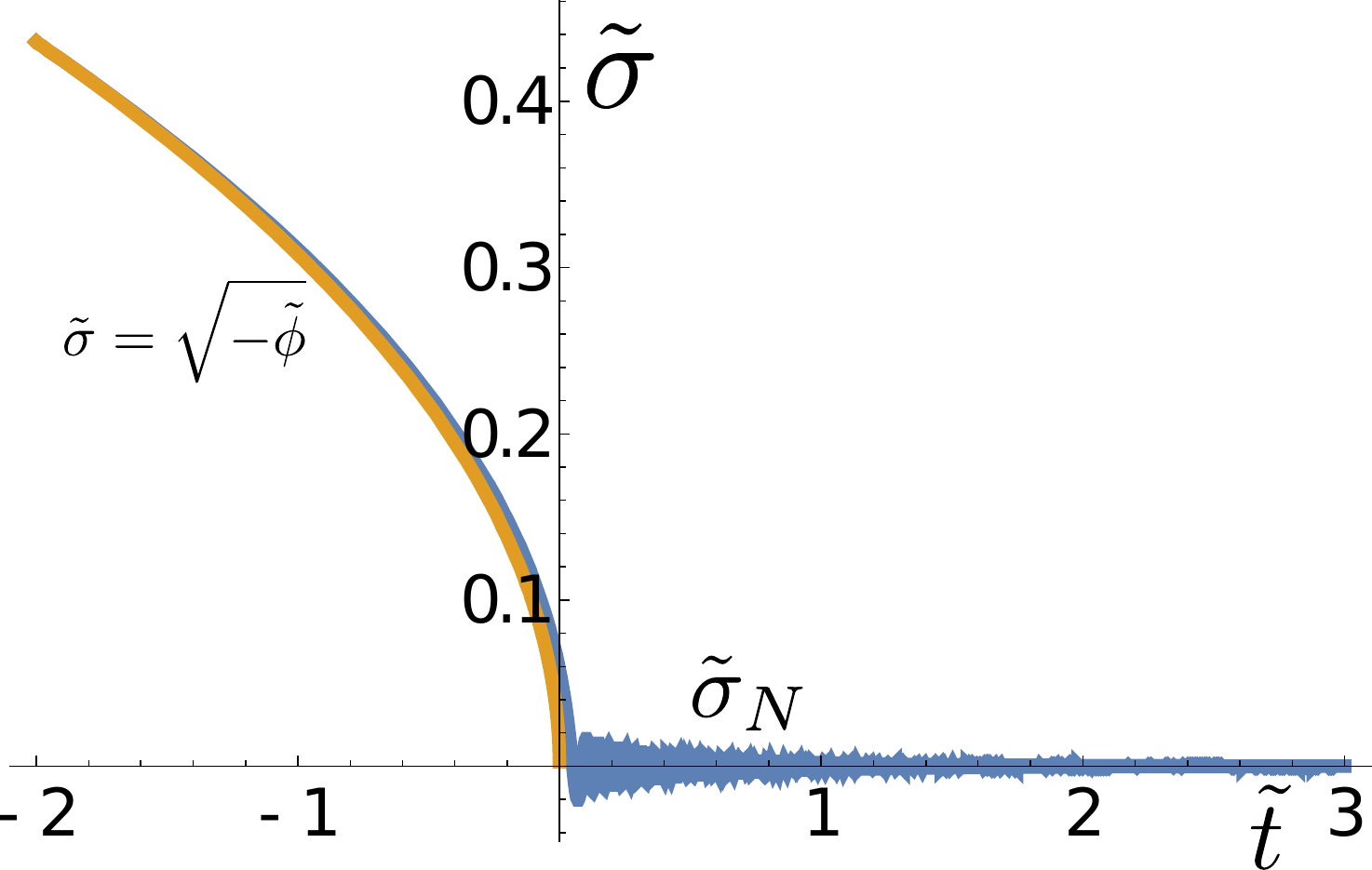}
\hspace{.1cm}
\includegraphics[width=0.31\textwidth]{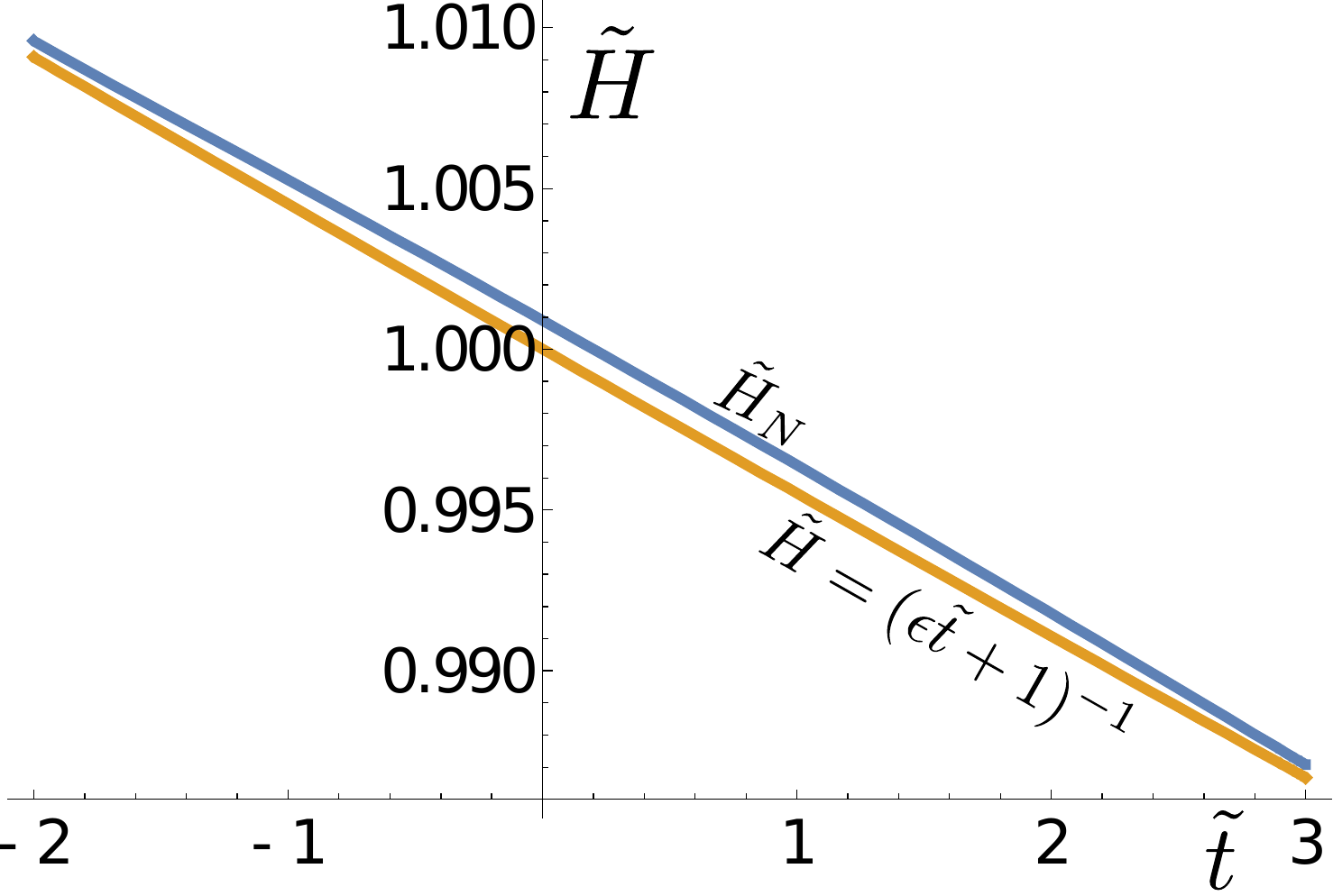}
\caption{Plots of $\tilde{\phi}$ (left), $\tilde{\sigma}$, (middle), and $\tilde{H}$ (right) for both the numerical and expected solutions. All three plots show that the analytical solutions are very good approximations to the numerical ones.}
\label{fig:numsol}
\end{figure}

We show in Figure~\ref{fig:numsol} the numerical solutions of equation~\ref{eq:rescaled_background}, which we label $\tilde{\phi}_N$, $\tilde{\sigma}_N$, and $\tilde{H}_N$, along with the analytical approximations: $\tilde{\phi} = \sqrt{2 \epsilon} \tilde{t}$, $\tilde{\sigma} = \sqrt{- \sqrt{2 \epsilon} \tilde{t}}$, and $\tilde{H} = 1/(\epsilon \tilde{t} + 1)$ where $\tilde{t} = \bar H (t - t_*)$. The constants chosen for the numerical solutions are
\begin{align}
& \epsilon = 0.0045 \, , & \eta = \frac{n_s -1 + 6 \epsilon}{2} \simeq - 0.0065 \, , \nonumber \\ & g_\phi =6 |\eta| \simeq 0.04 \, , & g_\sigma =\Lambda_\sigma^3/(2\,\sqrt{2\,\epsilon}\,\bar{H}^3) \simeq 2 \times 10^6 \, ,
\end{align}
where we have fixed $\epsilon\simeq .07/16$  by imposing that the ``vacuum'' tensor spectrum has maximum amplitude, while $\eta$ is determined by setting the spectral index $n_s=.96$. Finally, $g_\phi$ and $g_\sigma$ are determined by saturating the inequalities that appear below in Section~\ref{sec:gwlarge}. Even if for this choice of parameters the inequalities of Section~\ref{sec:gwlarge} are fully saturated, Figure~\ref{fig:numsol} shows that the analytical approximation for the evolution of the zero modes $\sigma_0$ and $\varphi_0$  does provide an excellent approximation of the exact evolution of the system.

\subsection{Scalar perturbations before the event of $\chi$ production}

We require that the metric perturbations are simply given by the usual single field formula $\zeta=-H\,\delta\varphi/\dot\varphi_0$. 

Well before the event of particle production, that is for $t\ll t_*-{\rm {Max}}\left\{\Lambda_\chi^{-1},\,\Lambda_\sigma^{-1}\right\}$ our system is in general described by two fields, $\varphi$ and $\sigma$, since $\chi$ is vanishing and irrelevant at this stage. The general expression for the curvature perturbation is 
\begin{align}
\zeta=-H\,\frac{\delta\rho}{\dot\rho}=-H\,\frac{\dot\varphi_0\,\delta\dot\varphi+\dot\sigma_0\,\delta\dot\sigma+ \left[V'(\varphi_0)+ \frac{\mu}{2}\,\sigma_0^2\right]\,\delta\varphi+\left[\mu\,\sigma_0\,\varphi_0+\lambda\,\sigma_0^3\right]\,\delta\sigma}{-3\,H\,\dot\varphi_0^2-3\,H\,\dot\sigma_0^2} \, .
\end{align}
Since the fluctuations of the field $\sigma$ are heavy at those times, $m_\sigma^2\simeq \Lambda_\sigma^3\,\left(t_*-t\right)\gg \Lambda_\sigma^2\gg H^2$, we can neglect $\delta\sigma$ in the equation above. Moreover, since the fluctuations of the field $\varphi$ become constant in the super horizon limit, we can neglect the term in $\delta\dot\varphi$. As a consequence, the expression for $\zeta$ simplifies to 
\begin{align}
\zeta=-H\,\frac{\left[V'(\varphi_0)+ \frac{\mu}{2} \,\sigma_0^2\right]\,\delta\varphi}{-3\,H\,\dot\varphi_0^2-3\,H\,\dot\sigma_0^2} \, .
\end{align}
In order to simplify this expression to its standard single field form we will then impose the following two requirements
\begin{align}
&(i)\qquad  \frac{\mu}{2}\,\sigma_0^2\ll |V'(\varphi_0)|\,,\nonumber\\
&(ii)\qquad |\dot\sigma_0|\ll |\dot\varphi_0|\,.
\end{align}
It is straightforward to see that condition {\em (i)} is equivalent to the requirement that that term in $\sigma_0$ in eq.~(\ref{eq:rescaled_background}) be negligible. This implies that condition {\em (i)} is satisfied whenever  eq.~(\ref{eq:gphismall}) holds.  

Condition {\em (ii)} is equivalent to 
\begin{align}
\frac{\mu}{4 \lambda} \ll \dot \varphi_* | t - t_*| \, ,
\end{align}
which we want to be satisfied for at least $|t-t_*| \gtrsim \Lambda_\sigma^{-1}$, leading to
\begin{align}
\frac{\mu^4}{2^5 \lambda^3} \ll 2\,\epsilon\,H^2\,M_P^2 \, .
\end{align}
%

\subsection{Energy density in the fluctuations of $\sigma$}
\label{sec:energysigma}
%

Since the fluctuations of $\sigma$ are sourced by the inflaton, energy conservation requires the energy density in those fluctuations to be smaller that the inflaton's kinetic energy.  We can compute the number density of $\sigma$ using the Bogolyubov coefficients computed in Subsection \ref{subsec:prodsigma} above, obtaining
\begin{equation}
n_\sigma = \int \frac{d^3 \textbf{p}}{a^3\,(2\pi)^3} |\beta_p|^2 \simeq { 1.7 \times 10^{-3} }\, \Lambda_\sigma^3\, \frac{a_*^3}{a^3}\,,
\end{equation}
while the energy density reads approximately
\begin{equation}
\rho_\sigma = n_\sigma\, m_\sigma ={ 1.7 \times 10^{-3}} \,\Lambda_\sigma^3\,\frac{a_*^3}{a^3} \, \sqrt{\mu \varphi}\,.
\end{equation}

Approximating $\varphi\simeq \dot\varphi_*\,(t-t_*)$ and $a_*/a\simeq e^{-H(t-t_*)}$, we see that $\rho_\sigma$ is maximized for $t-t_*\simeq (6\,H)^{-1}$, where it evaluates to
\begin{equation}
\rho_\sigma^{\rm {max}} \simeq{ 6.0 \times 10^{-4}} \,\Lambda_\sigma^3\,\sqrt{\frac{\mu\,\dot\varphi_*}{H}}\,,
\end{equation}
that we require to be smaller than the kinetic energy of the inflaton $\dot\varphi_*^2/2$, leading to the constraint
\begin{align}
\frac{\mu^3}{\sqrt{2\epsilon}\, H^2\,M_P}\ll 1.7 \times 10^5 \, .
\end{align}

\subsection{Energy density in the fluctuations of $\chi$}

By an argument analogous to that of the previous subsection we require the energy in the $\chi$ particles to be smaller than the kinetic energy in the zero mode of the $\sigma$ field.

Inserting eqs.~(\ref{eomml}) and~(\ref{cp1}) into the expression for the energy in modes of $\chi$,
\begin{equation}
\braket{\rho_\chi} = \frac{1}{2\,a^4} \int \frac{d^3 \textbf{p}}{(2\pi)^3}\left[ \chi'_p {\chi_p^*}' + \chi_p \chi_p^*\left( p^2 - \frac{2}{\tau^2} \right)  \right],
\end{equation}
we obtain an expression that is ultraviolet divergent. To make it finite we subtract off the energy in the mode functions computed for $\Lambda_\chi=0$,
\begin{equation}
\braket{\rho_\chi}\to \frac{1}{2\,a^4} \int \frac{d^3 \textbf{p}}{(2\pi)^3}\left[ \left( \chi'_p \,{\chi_p^*}{}' - \tilde{\chi}'_p \,{\tilde{\chi}_p^*}{}'\right)+ \left( \chi_p\, \chi_p^* - \tilde{\chi}_p\, \tilde{\chi}_p^* \right)\left( p^2 - \frac{2}{\tau^2} \right)  \right],
\end{equation}
where the functions $\chi_p$ and $\tilde{\chi}_p$ are given in eqs.~(\ref{eomml}) and~(\ref{adiabaticchi}) above.  Numerical integration then gives the result
\begin{equation}
\braket{\rho_\chi} = 8 \times 10^{-4} \Lambda_\chi^4\, \frac{a_*^4}{a^4}\,,
\end{equation}
that is maximal when $a=a_*$. 

Since quanta of $\chi$ are produced by the rolling of the field $\sigma$, energy conservation requires $\braket{\rho_\chi}\ll \frac{1}{2}\dot\sigma(t_{\rm {prod}})^2$ where $t_{\rm {prod}}$ is the time at which the production of most of the quanta of $\chi$ occurs. We note that, for $t<t_*$, $\sigma(t)\sim \sqrt{\frac{\mu}{\lambda}\,\dot\varphi_*\left(t_*-t\right)}$, so that $\dot\sigma$ is divergent as $t\to t_*$. However, the production happens at a typical time of the order $t\simeq t_*+{\cal O}(\Lambda_\chi^{-1})$. As a consequence, we will impose $\braket{\rho_\chi}\ll \frac{1}{2}\dot\sigma(t_*+{\cal O}(\Lambda_\chi^{-1}))^2\sim \Lambda_\chi^4/h^2$, leading to the constraint $h\ll 30$ that is always satisfied since we  require $h\lesssim 1$ by perturbativity.

\subsection{Effect of the fluctuations of $\sigma$ on the metric perturbations}
\label{sec:sigmaonphi}
%

We next consider how the fluctuations in $\sigma$ will affect the fluctuations in $\varphi$. In particular, we want to make sure that those induced fluctuations in $\varphi$ are small compared to the scalar perturbations measured in the CMB. Since the fluctuations of $\sigma$ are significant only after $t_*$  we set $\sigma_0 = 0$ so that we are left with the following equation for the fluctuations in $\varphi$:
\begin{align}
\delta\varphi''+2\,\frac{a'}{a}\delta\varphi'-\Delta\,\delta\varphi + \frac{\mu}{2}\,a^2\, \delta\sigma^2 = 0 \,,
\end{align}
which we solve as
\begin{equation}
\delta \varphi(\textbf{k},\tau) = \frac{\mu}{2} \int d\tau' a^2(\tau')\, G_k(\tau,\tau') \int \frac{d^3\textbf{p}}{(2\pi)^{3/2}}\, \delta \sigma(\textbf{p},\tau')\, \delta \sigma(\textbf{k}-\textbf{p},\tau') \, ,
\end{equation}
where the Green's function reads
\begin{equation}
G_k(\tau,\tau') = \frac{1}{k^3  {\tau'}^2} \left[ (1 + k^2 \tau \tau')\, \sin (k (\tau - \tau')) + k\,(\tau' - \tau)\, \cos (k(\tau-\tau')) \right] \Theta(\tau - \tau').
\end{equation}

We are ultimately interested in the power spectrum for the fluctuations in $\varphi$ so we first calculate the correlator,
\begin{align}\label{twoptphi}
\langle\delta\varphi(\bk,\,\tau)\,\delta\varphi(\bk',\,\tau)\rangle&=\frac{\mu^2}{4}\int\frac{d\tau'}{H^2\tau'{}^2}\frac{d\tau''}{H^2\tau''{}^2}\,G_k(\tau,\,\tau')\,G_{k'}(\tau,\,\tau'') \nonumber\\
&\times\int\frac{d\bp\,d\bp'}{(2\pi)^3}\langle\delta\sigma(\bp,\,\tau')\,\delta\sigma(\bk-\bp,\,\tau')\delta\sigma(\bp',\,\tau'')\,\delta\sigma(\bk'-\bp',\,\tau'')\rangle\,.
\end{align}

The correlator for $\sigma$ is given by an equation analogous to eq.~(\ref{twoptchi1}), with the modes for $\delta \sigma$ quickly becoming nonrelativistic after $t_*$ since $m_\sigma\sim \sqrt{\mu\,\dot\varphi_*\,(t-t_*)}$ continues to grow. As a consequence, dropping the terms that are quickly oscillating, we obtain
\begin{align}
&\int d\bp\,d\bp'\braket{ \delta \sigma (\textbf{p}, \tau') \delta \sigma (\textbf{p}- \textbf{k}, \tau') \delta \sigma (\textbf{p}', \tau'') \delta \sigma (\textbf{p}'- \textbf{k}', \tau'') }\simeq \frac{\delta^{3}(\textbf{k}+\textbf{k}')}{2\, a^2(\tau')\,a^2(\tau'')\, \omega_\sigma(\tau')\,\omega_\sigma(\tau'')} \nonumber\\
&\times\int d\bp \, \left\{ |\beta_\sigma(p)|^2\, |\beta_\sigma(|{\textbf{k}-\textbf{p}}|)|^2 + \mathrm{Re}\left[ \alpha_\sigma(p)\, \beta^*_\sigma(p) \,\alpha^*_\sigma(|{\textbf{k}-\textbf{p}}|) \,\beta_\sigma(|{\textbf{k}-\textbf{p}}|) \right] \right\} \, .
\end{align}

Also, we will compute the correlator for modes that are well outside of the horizon at the end of inflation, so that we can set $\tau\to 0$ in eq.~(\ref{twoptphi}). Collecting everything we get
\begin{align}\label{twoptphi1}
\langle\delta\varphi(\bk,\,\tau)\,\delta\varphi(\bk',\,\tau)\rangle& = \frac{\mu^2 H^3 \delta^{(3)}(\textbf{k} + \textbf{k}')}{32 \pi^3 k^3 \,\Lambda_\sigma^3}\frac{1}{k^3}\left[\int_{\tau_*}^0\frac{d\tau'}{-\tau'}\,\frac{\sin k\tau'-k\tau'\,\cos k\tau'}{\sqrt{\ln\left(\frac{\tau_*}{\tau'}\right)}}\right]^2\nonumber\\
&\times\int d\bp \left\{ |\beta_\sigma(p)|^2 |\beta_\sigma(|{\textbf{k}-\textbf{p}}|)|^2 + \mathrm{Re}\left[ \alpha_\sigma(p) \beta^*_\sigma(p) \alpha^*_\sigma(|{\textbf{k}-\textbf{p}}|) \beta_\sigma(|{\textbf{k}-\textbf{p}}|) \right] \right\}\,.
\end{align}

To calculate the momentum integral, we note that the integral in $d\bp$ gets most of its contributions by $p={\cal O}(\Lambda_\sigma a_*)$. On the other hand, the temporal function of $\tau_*$ forces $k={\cal O}(|\tau_*|^{-1})\ll {\cal O}(\Lambda_\sigma a_*)$. Therefore we can neglect the $k$-dependence inside the momentum integral, so that the second line of eq.~(\ref{twoptphi1}) can approximated by
\begin{align}
\int d\bp  \left\{ 2\, |\beta_\sigma(p)|^4 + |\beta_\sigma(p)|^2 \right\}\simeq .51\,a_*^3\,\Lambda_\sigma^3\,.
\end{align}

\begin{figure}
\centering
\includegraphics[width=0.5\textwidth]{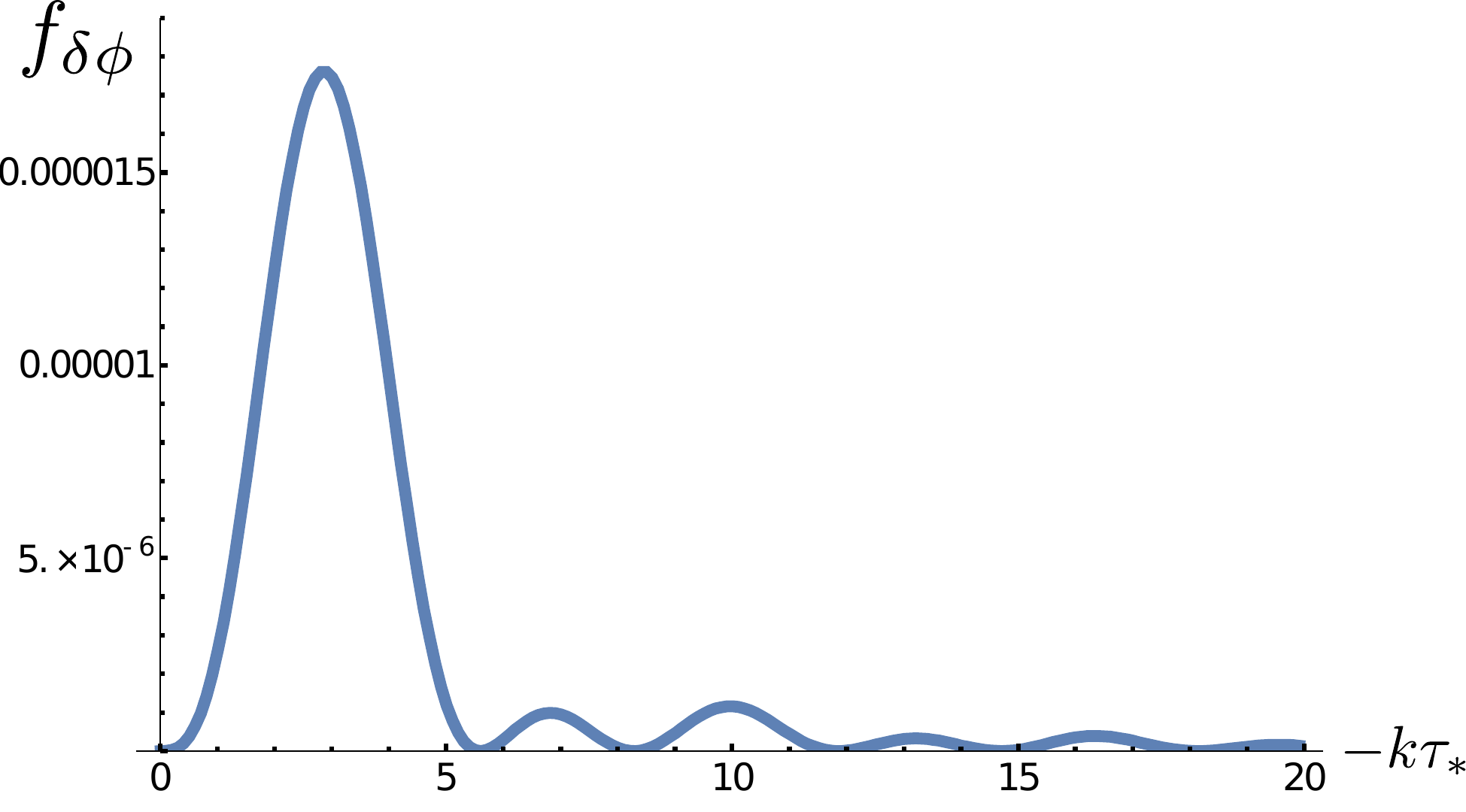}
\caption{Plot of the function $f_{\delta\varphi}(-k\tau_*)$ defined in eq.~(\ref{psdeltaphi}).}\label{fig:deltaphi}
\end{figure}

Thus we finally find the power spectrum of fluctuations in $\delta\varphi$ induced by the fluctuations in $\delta\sigma$
\begin{align}\label{eq:sourced_spectrum}
{\cal P}_{\delta\varphi}^{\rm sourced}=\mu^2\,f(-k\,\tau_*)\,,
\end{align}
where
\begin{align}\label{psdeltaphi}
f(y)\simeq \frac{.51}{64 \pi^5 y^3}  \left[\int\limits^{y}_0 \frac{dx}{x}\,\frac{ - \sin(x) + x \,\cos(x)}{\sqrt{\ln\left(\frac{y}{x}\right)}}\right]^2 \, ,
\end{align}
is plotted in Figure~\ref{fig:deltaphi} and is maximized at $y\simeq 3$ where it evaluates to $1.7\times 10^{-5}$.

To sum up, the requirement that the metric perturbations induced by the interactions between the inflaton and the field $\sigma$ do not exceed the measured amplitude of the scalar power spectrum leads to the constraint
\begin{align}
\frac{\dot \varphi_*^2}{H^2}\, {\cal P}_\zeta \simeq \left( \frac{H}{2\pi} \right)^2 \gg {\cal P}_{\delta \varphi}^{\rm sourced}(k) \hspace{0.5cm} \Rightarrow \hspace{0.5cm} 25 \times 10^{-10} \frac{\dot \varphi_*^2}{H^2}\gg  \mu^2   f(-k\tau_*)  \,,
\end{align}
or, equivalently,
\begin{align}\label{eq:constr_pzeta}
\frac{\mu^2}{2\epsilon\,M_P^2}\ll 1.5 \times 10^{-4} \, .
\end{align}

Before concluding this Section, we note that the sourced scalar perturbations will obey non-gaussian statistics and would be in principle subject to the strong constraints from the Planck satellite on the amplitude of equilateral bispectra~\cite{Ade:2015ava}. However, those strong constraints hold when the nongaussian component of the scalar perturbations has a (quasi) scale invariant component. Since the contribution~(\ref{eq:sourced_spectrum}) is strongly scale dependent, we do not expect the non-observation of equilateral nongaussianities to constrain the parameter space of the model more strongly than eq.~(\ref{eq:constr_pzeta}), similarly to what was found in~\cite{Namba:2015gja}.

\section{How large of a spectrum for induced PGWs can be generated?}
\label{sec:gwlarge}
%

The spectrum of produced gravitational waves is proportional to $\Lambda_\chi^5$, and in this Section we estimate how large $\Lambda_\chi$ can be once the constraints of the previous section are enforced.  The constraints found in the previous section can be summarized as follows
\begin{align}\label{eq:initial_inequalities}
&(i)\qquad 2\mu\,\sqrt{2\epsilon}\,M_P\gg H^2 & {\rm Subsection\ 4.2}\, , \nonumber\\
&(ii)\qquad \frac{h^2\,\mu\,\sqrt{2\epsilon}\,M_P}{\lambda}\gg H^2 &{\rm Subsection\ 4.2}\, , \nonumber\\
&(iii)\qquad \frac{\mu^2}{\lambda\,H^2}\ll 6\,|\eta|  & {\rm Subsection\ 4.3}\, ,\nonumber\\
&(iv)\qquad \frac{\mu^4}{2^5 \lambda^3} \ll 2\,\epsilon\,H^2\,M_P^2 & {\rm Subsection\ 4.4}\, , \nonumber\\
&(v)\qquad \frac{\mu^3}{\sqrt{2\epsilon}\, H^2\,M_P}\ll 1.7 \times 10^5 & {\rm Subsection\ 4.5}\, ,\nonumber\\
&(vi)\qquad \frac{\mu^2}{2\epsilon\,M_P^2}\ll 1.5 \times 10^{-4} & {\rm Subsection\ 4.7}\, ,
\end{align}
besides the perturbativity requirements $h,\,\lambda<1$. 

\subsection{Detectability at CMB scales}

\begin{figure}
\centering
\includegraphics[width=0.6\textwidth]{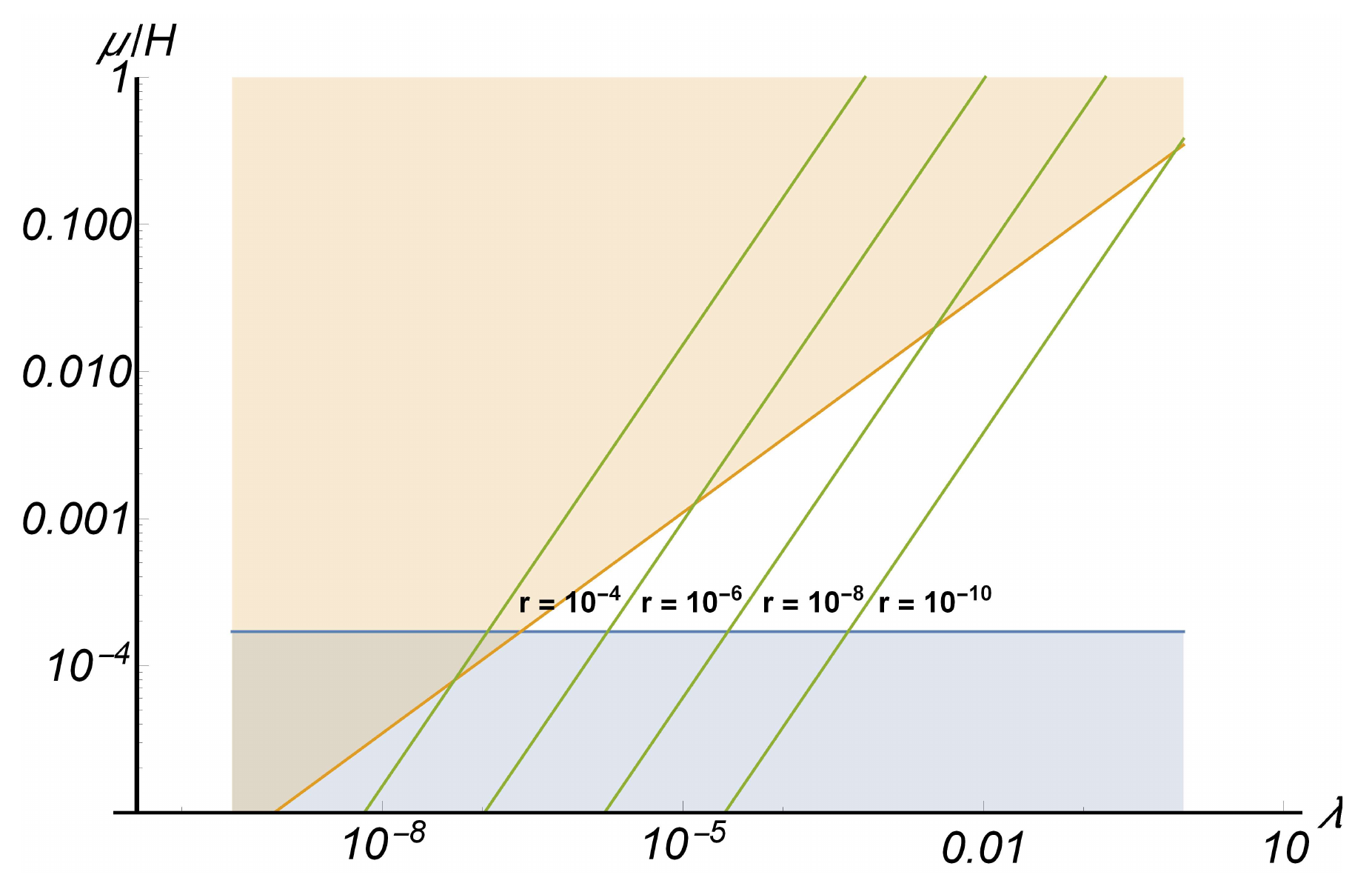}
\caption{Constraint plot for equation~(\ref{mulcons}) showing the allowed parameter space for $\mu$ and $\lambda$. Lines of constant $r$ are showed ranging from $r=10^{-4} - 10^{-10}$.}\label{fig:mulcons}
\end{figure}

Let us first consider the maximal amplitude of the sourced gravitational  waves at CMB scales, where the dynamics of the inflaton is constrained by CMB observations. First, we trade $M_P$ for $H$ and $\epsilon$ using COBE normalization $2\epsilon\simeq 10^7\,H^2/M_P^2$:
\begin{align}
&(i)\qquad \mu\gg 1.7\times 10^{-4}\,H\, , \nonumber\\
&(ii)\qquad \frac{h^2}{\lambda}\mu\gg 1.7\times 10^{-4}\,H\, \, , \nonumber\\
&(iii)\qquad \frac{\mu^2}{\lambda}\ll 6\,|\eta|\,H^2\, ,\nonumber\\
&(iv)\qquad \frac{\mu^4}{\lambda^3} \ll 3\times 10^8\,H^4 \, , \nonumber\\
&(v)\qquad {\mu^3}\ll  5\times 10^8\,H^3 \, ,\nonumber\\
&(vi)\qquad {\mu^2}\ll 1.5 \times 10^{3}\,H^2\, .\nonumber\\
\end{align}
We now note, first, that $h$ appears only in condition {\em (ii)}. In order to maximize the volume of our parameter space while remaining within the perturbative regime we set from now on $h=1$. Moreover, we note that, once the perturbativity requirement $\lambda<1$ is imposed, conditions {\em (i)}, {\em (iii)}, {\em (iv)}, {\em (v)} and {\em (vi)} reduce simply to conditions {\em (i)} and {\em (iii)}. Therefore we are left just with
\begin{align}
&(i)\qquad \mu\gg 1.7\times 10^{-4}\,H\, , \nonumber\\
&(iii)\qquad \frac{\mu^2}{\lambda}\ll 6\,|\eta|\,H^2\,.
\end{align}
We remember that we are seeking to maximize $\Lambda_\chi/H$, which is proportional to $(\mu/(\lambda\,H))^{1/3}$. Trading $\mu$ for $\Lambda_\chi$ in the equations above by using
\begin{align}
\Lambda_\chi^3=\frac{h^2\,\mu}{\lambda}\,\sqrt{2\epsilon}\,H\,M_P\to 3\times 10^3\,\frac{\mu}{\lambda}\,H^2\,,
\end{align}
we obtain the following constraints
\begin{align} \label{mulcons}
&(i)\qquad \lambda\,\Lambda_\chi^3\gg .5\,H^3\, , \nonumber\\
&(iii)\qquad \lambda^{1/2}\,\Lambda_\chi^3\ll 8\times 10^3\,\sqrt{|\eta|}\,H^3\, ,
\end{align}
where $\Lambda_\chi$ is maximized by setting 
\begin{align}
\lambda\simeq 4\times 10^{-9}\,|\eta|^{-1},\qquad \Lambda_\chi\simeq 500\,|\eta|^{1/3}\,H\,.
\end{align}
From now on we set $|\eta|=.02$ to fix ideas (this is the value one obtains if one assumes that $\epsilon$ gives a negligible contribution to the scalar spectral index $n_s=1+2\,\eta-6\,\epsilon\simeq .96$). Then, trading $\mu$ for $r_{\rm sourced}=10^3\,\Lambda_\chi^5/(H\,M_P^4)$ and for $r_{\rm vacuum}=.8\times 10^8\,H^2/M_P^2$, so that
\begin{align}\label{upper_rind}
\frac{r_{\rm sourced}}{r_{\rm vacuum}}\simeq \left(\frac{r_{\rm vacuum}}{0.07}\right)\,\left(\frac{\Lambda_\chi}{620\,H}\right)^5\ll 5\times 10^{-4}\,\left(\frac{r_{\rm vacuum}}{0.07}\right)\,.
\end{align}
We conclude therefore that the sourced component, in the case of a single $\chi$ species, can give at most a ${\cal O}(.1\%)$ contribution to the vacuum contribution to the primordial spectrum of tensors, and that such a situation  is obtained in the regime where the vacuum contribution is maximal while in agreement with the existing observational constraints. Figure \ref{fig:mulcons} shows the constraint plot for the allowed value of $\mu$ and $\lambda$ with constant lines $r$ using the above relations.  

We finally note that Figure~\ref{fig:numsol} shows an excellent agreement between the analytical approximation and the actual numerical solution to the background evolution equations. In that Figure, the constraint {\em (iii)} above, which limits the amplitude of the sourced tensors, is fully saturated. Therefore, it is possible that the constraint {\em (iii)} might even be violated by a factor $10$  or so without changing significantly the dynamics of the system. This in turn implies that the bound~(\ref{upper_rind}) might be slightly too restrictive. We do not expect, however, this consideration to significantly affect our conclusion that the sourced component is well subdominant with respect to the vacuum one, at least in the case of a single (or a few) $\chi$ species.

\subsection{Direct detectability by gravitational wave interferometers}

Next, we ask the question of whether it would be possible to obtain tensors with a larger amplitude at smaller scales directly probed by gravitational interferometers such as Advanced LIGO or LISA \cite{Harry:2010zz,Bartolo:2016ami}. While, on the one hand, the sensitivity of those experiments to primordial gravitational waves is much weaker than that of CMB polarization, on the other hand the system is not subject to the constraints imposed by CMB observations. In the case of production of primordial gravitational waves by the amplification of vacuum fluctuations of gauge fields, this allows for observable gravitational waves at interferometer scales \cite{Cook:2011hg}. 

In the case of the present model, however, the upper bound~(\ref{upper_rind}) is imposed by conditions {\em (i)} and {\em (iii)} above, which in turn derive just from the requirement of the consistency of the background dynamics, and do not depend strongly on the CMB constraints. The only difference is that we can disentangle $\epsilon$ and $H$ by not imposing the COBE relation $2\epsilon\simeq 10^7\,H^2/M_P^2$. More explicitly, by imposing $h=1$, we obtain
\begin{align}
\Omega_{\rm GW}\,h^2\ll 1.2\times 10^{-11}\,\lambda^{-5/3}\,\frac{\mu^{5/3}H^{2/3}}{M_P^{7/3}}\,.
\end{align}
Now, all inequalities~(\ref{eq:initial_inequalities}) are best satisfied when  $\epsilon$ and $|\eta|$ are largest. When both slow roll parameters are of the order of the unity, the most stringent among those inequalities are again {\em (i)} and {\em (iii)}. If we saturate them we obtain
\begin{align}
\Omega_{\rm GW}\,h^2\ll 1.4\times 10^{-9}\,\left(\epsilon\,|\eta|\right)^{5/3}\,\left(\frac{H}{M_P}\right)^{2/3}\,.
\end{align}
Finally, we note that energy conditions require that the value of the Hubble parameter at the smaller scales probed by CMB interferometers must be  smaller than the value of the same quantity at CMB scales, which is constrained by observations to $H=1.1\times 10^{-4}\,\sqrt{r_{\rm vacuum}}\,M_P<3\times 10^{-5}\,M_P$. As a consequence we get the absolute upper bound
\begin{align}
\Omega_{\rm GW}\,h^2\ll 1.3\times 10^{-12}\,\left(\epsilon\,|\eta|\right)^{5/3}\,.
\end{align}

While this figure, for large values of $\eta,\,\epsilon={\cal O}(1)$, is above the projected sensitivity of LISA, $\Omega_{\rm GW}\,h^2\simeq 10^{-13}$ in its most optimistic configuration, we should stress that it has been obtained by saturating a  few ``much larger'' inequalities: $X\gg Y\to X=Y$ and that the ``natural'' value of the slow roll parameters, at these scales that exit the horizon closer to the end of inflation, is of the order of $\sim 10^{-1}$. Therefore we are led to the conclusion that our scenario will generally not lead to a detectable effect at the LISA level. It is worth stressing, however, that a  feature in the inflationary potential leading to larger values of $\epsilon$ and $\eta$, along with the presence of a number $N_\chi>1$ of $\chi$ species that will enhance our effect by a factor $N_\chi$, might be able to bring it into the observable window without requiring a huge stretch of parameters.

\section{Conclusions}
\label{sec:conclusion}
%

Production of inflationary gravitational waves by resonant production of scalars, first studied in~\cite{Cook:2011hg} (see also~\cite{Senatore:2011sp,Barnaby:2012xt,Carney:2012pk}) is known to be inefficient. The main cause of such inefficiency is attributed~\cite{Barnaby:2012xt} to the fact that, in the model of~\cite{Cook:2011hg}, the scalar that sources gravitational waves becomes very massive soon after the event of particle production. This observation has motivated us to study a similar mechanism in the case of a system that undergoes symmetry restoration: a scalar $\chi$, whose mass is controlled by an order parameter $\sigma$ suddenly goes from massive to massless. 

We have found that the amplitude for the spectrum of gravitational waves obtained in this scenario is indeed larger by some orders of magnitude than that of~\cite{Cook:2011hg}. More specifically, if we impose CMB constraints, the sourced component of tensors can yield a sourced tensor-to-scalar ratio as large as $r_{\rm sourced}\sim 10^{-5}$ per $\chi$ species (in~\cite{Cook:2011hg} the corresponding figure was $\sim 10^{-8}$). Since the largest value of $r_{\rm sourced}\sim 10^{-5}$ is obtained under the assumption that the vacuum contribution to the tensor spectrum is the largest one compatible with current observations, $r_{\rm vaucuum}=.07$, we expect that detection of the sourced component, even in the most optimistic scenario, and possibly assuming a ${\cal O}(10)$ boost factor to account for multiple $\chi$ species, will be extremely challenging. The main signature of the sourced component would be an oscillating feature on the top of the spectrum of the smooth vacuum component.

If we relax the constraints from CMB and simply require that the inflaton is still in slow roll ($\epsilon,\,|\eta|<1$) then we can get a contribution to the energy density on gravitational waves $\Omega_{GW}\,h^2$ that in our case can be as large as $10^{-12}$ per $\chi$ species (in~\cite{Cook:2011hg} the corresponding figure was $\sim 10^{-20}$). This situation is relevant for smaller scales, where the constraints from CMB do not hold, and that would be of interest for gravitational interferometers. For reference, the projected sensitivity of LISA is $\Omega_{GW}\,h^2\simeq 10^{-13}$~\cite{Bartolo:2016ami}. We stress, however, that the maximum amplitude $\Omega_{GW}\,h^2\sim 10^{-12}$ is obtained in our model by looking at a very narrow, and unlikely even if not forbidden, portion of parameter space. It would be interesting to perform a more detailed, possibly fully numerical, analysis of the resulting spectrum in a concrete model of inflation. 

Finally, we note that if the symmetry that gets restored is a gauge symmetry, then the mechanism discussed in this paper would lead to the generation of gauge bosons. In~\cite{Barnaby:2012xt} it was shown that, for models where the produced particle becomes massive shortly after production, the spectrum of  gravitational waves sourced by vectors had the same amplitude as that sourced by scalars (times a factor that accounted for the different number of degrees of freedom).  It is not obvious, however, that such a result would hold also for vectors. Moreover, one of the two most constraining factors {\em (i)} and {\em (iii)} of Section~\ref{sec:constraints} that limits the amplitude of the sourced tensor component does emerge from the requirement of the consistency of the background dynamics for this very specific model. It would be interesting to study whether other mechanisms that lead to symmetry restoration during inflation could give different results. We plan to attack these questions in future work.

\smallskip

{\bf Acknowledgments.} We thank Valerie Domcke for useful discussions. This work has been partially supported by the NSF grant PHY-1520292.

\appendix

\section{Gravitational Wave Production in Minkowski Space}
\label{app:min}
%

Since gravitational waves are expected to be produced at a typical frequency $\Lambda_\chi\gg H$, we can extract most information about this process by working on a Minkowski background. Such an approach simplifies the task of computing finite quantities, since on a Minkowski background all the modes are evolving adiabatically after the event of production of quanta of $\chi$ and one can fully use the machinery provided by the Bogolyubov coefficients.

\subsection{Production of quanta of $\chi$}

Let us first characterize the production of quanta of the field $\chi$ induced by the time-dependence~(\ref{chimass}) of its mass. We decompose the field $\chi$  as
\begin{equation}
\hat \chi(\textbf{x},t) = \int \frac{d^3 \textbf{p}}{(2 \pi)^{3/2}} e^{i \textbf{p} \cdot \textbf{x}} \left[ \chi_\textbf{p}(t) \,\hat a_\textbf{p} + \chi^*_\textbf{-p}(t)\, \hat a^\dagger_\textbf{-p} \right]\,,
\end{equation}
where the mode functions satisfy
\begin{align}\label{eomlinear}
&\ddot \chi_\textbf{p} + \left[ p^2 -\Lambda_\chi^3\,t\right]\, \chi_\textbf{p} = 0\,,\qquad t<0\,,\nonumber\\
&\ddot \chi_\textbf{p} + p^2\, \chi_\textbf{p} = 0\,,\qquad t>0\,,
\end{align}
whose solution  for $t < 0$ can be written as
\begin{equation}
\chi_p(t < 0) = \sqrt{\frac{\pi\, z}{6\,\Lambda}} \mathrm{H}^{(1)}_{\frac{1}{3}}\left( \frac{2}{3} z^{\frac{3}{2}}  \right)\,, \hspace{1.0cm} z \equiv \frac{p^2}{\Lambda_\chi^2} - \Lambda_\chi\, t.
\end{equation}
where $H^{(1)}_\nu(z)$ denotes the Hankel function of the first kind and where we have determined the integration constants assuming that the modes of $\chi$ are in their adiabatic vacuum at early times, 
\begin{equation}
\chi^{\text{WKB}}_p(t\to-\infty) \to \frac{e^{-i \int \omega(t')\, dt'}}{\sqrt{2\, \omega_p}} =\frac{1}{\sqrt{2\,\Lambda_\chi\,\sqrt{z}}} e^{i \frac{2}{3} z^{\frac{3}{2}}}\,.
\end{equation}

The solution of eq.~(\ref{eomlinear}) for $t > 0$ is simply the massless Klein-Gordon solution, which we write as
\begin{equation}
\chi_p(t > 0) = \frac{\alpha_p}{\sqrt{2\,p}} e^{- i\,p\, t} + \frac{\beta_p}{\sqrt{2\,p}} e^{i \,p\, t}\,,
\end{equation}
where the parameters $\alpha_p$ and $\beta_p$ are the Bogolyubov coefficients, which are determined by matching the mode functions and their first derivatives at time $t = 0$, and are given by
\begin{eqnarray}
&&\alpha_p = \sqrt{\frac{\pi }{12}}\, \left(\frac{p}{\Lambda_\chi}\right)^{3/2}\, \left(H_{\frac{1}{3}}^{(1)}\left(\frac{2\,p^3}{3\,\Lambda_\chi^3}\right)-i\,H_{-\frac{2}{3}}^{(1)}\left(\frac{2\,p^3}{3\,\Lambda_\chi^3}\right)\right)\,,\nonumber\\
&&\beta_p =  \sqrt{\frac{\pi }{12}}\,\left(\frac{p}{\Lambda_\chi}\right)^{3/2} \left(H_{\frac{1}{3}}^{(1)}\left(\frac{2\,p^3}{3\,\Lambda_\chi^3}\right)+i\,H_{-\frac{2}{3}}^{(1)}\left(\frac{2\,p^3}{3\,\Lambda_\chi^3}\right)\right).
\end{eqnarray}

The Bogolyubov coefficient $\beta_p$ has the following behavior 
\begin{align}
&\beta_p\simeq \frac{\sqrt{\pi}\left(-1+i\,\sqrt{3}\right)}{2\times 3^{1/3}\,\Gamma(1/3)}\,\sqrt{\frac{\Lambda_\chi}{p}}+{\cal O}\left(\sqrt{\frac{p}{\Lambda_\chi}}\right)\simeq (-.229+.397\,i)\sqrt{\frac{\Lambda_\chi}{p}}+{\cal O}\left(\sqrt{\frac{p}{\Lambda_\chi}}\right)\,,\qquad p\to 0 \, ,\nonumber\\
&\beta_p\simeq (-.121-0.032\,i)\,e^{\frac{2}{3}\,i\,\frac{p^3}{\Lambda_\chi^3}}\,\frac{\Lambda_\chi^3}{p^3}+{\cal O}\left(\left(\frac{\Lambda_\chi}{p}\right)^4\right)\,,\qquad p\to \infty \, .
\end{align}
The number density for $\chi$ is then given by
\begin{equation}
n_\chi = \int \frac{d^3\textbf{p}}{(2\pi)^3} |\beta_p|^2 =\frac{\Lambda_\chi^3}{48 \sqrt{3} \pi^2},
\end{equation}
whereas the energy density is 
\begin{equation}
\rho_\chi = \int \frac{d^3\textbf{p}}{(2\pi)^3}\,p\, |\beta_p|^2 \simeq 8.09\times 10^{-4}\,\Lambda_\chi^4\,.
\end{equation}
%

\subsection{Generation of gravitational waves}

The equation of motion for the graviton in Minkowski space reads
\begin{equation}
\ddot{h}_{ij} - \Delta h_{ij} = \frac{2}{M_P^2} \Pi_{ij}^{\; \; ab}\,T_{ab}\,,
\end{equation}
where the spatial components of the stress energy tensor, expanding to first order in $h_{ab}$, read
\begin{equation}
T_{ab}=\partial_a\chi\,\partial_b\chi+h_{ab}\left[\frac{1}{2}\dot\chi^2-\frac{1}{2}\left(\nabla\chi\right)^2-V(\chi)\right]+\dots\,,
\end{equation}
where the dots denote terms that are second or higher order in $h_{ab}$ and terms that are proportional to $\delta_{ab}$ and are projected out by $\Pi_{ij}^{\; \; ab}$.

Next, we write $h_{ij}$ as $h_{ij}^{(0)}+h_{ij}^{(1)}$, where $h_{ij}^{(0)}$ is a solution of the homogeneous equation $\ddot{h}^{(0)}_{ij} - \Delta h_{ij}^{(0)} =0$. Therefore, to leading order, the equation for $h_{ij}^{(1)}$ is solved by
\begin{equation}\label{solmink}
h_{ij}^{(1)}({\bf p},\,t)= \frac{2}{M_P^2} \int dt' G_p(t,\,t') \,\Pi_{ij}^{\; \; ab}(\textbf{p}) \left(\partial_a\chi\,\partial_b\chi+h^{(0)}_{ab}\left[\frac{1}{2}\dot\chi^2-\frac{1}{2}\left(\nabla\chi\right)^2-V(\chi)\right]\right)\left({\bf p},\,t'\right)\, ,
\end{equation}
where the Green's function for the graviton on Minkowski space reads
\begin{equation}
G(t,t') = \frac{\sin (k( t - t'))}{k}\Theta(t-t')\,.
\end{equation}

Using the fact that $h_{ij}^{(0)}$ and $\chi$ are uncorrelated, we find the graviton correlator to leading order as
\begin{equation}
\braket{\left(h_{ij}^{(0)}(\textbf{k},t)+h_{ij}^{(1)}(\textbf{k}',t)\right)\,\left(h_{ij}^{(0)}(\textbf{k},t)+h_{ij}^{(1)}(\textbf{k}',t)\,\right)}=\frac{2\,\pi^2}{k^3}\delta(\textbf{k}+\textbf{k}')\,\left({\cal P}_T^{00}+{\cal P}_T^{11}+2\,{\mathrm Re}\left\{{\cal P}_T^{01}\right\}\right)\,,
\end{equation}
where $\frac{2\,\pi^2}{k^3}\delta(\textbf{k}+\textbf{k}'){\cal P}_T^{00}=\braket{h_{ij}^{(0)}(\textbf{k},t)  h_{ij}^{(0)}(\textbf{k}',t)}$ is the standard vacuum contribution that we will ignore in this section devoted to the production of gravitational waves in Minkowski space. Next we have
\begin{align} \label{P11mink}
&\frac{2\,\pi^2}{k^3}\,\delta(\textbf{k}+\textbf{k}')\,{\cal P}_T^{11} =  \frac{1}{2\, \pi^3\, M_P^4} \int dt'\, G_k(t,\,t') \int dt''\, G_{k'}(t,\,t'') \,\Pi_{ij}^{\; \; ab}(\textbf{k})\, \Pi_{ij}^{\; \; cd}(\textbf{k}') \nonumber \\ 
&\times \int d^3\textbf{p}\,d^3\textbf{p}'\, p_a\,(\textbf{k}_b - \textbf{p}_b)\, p'_c\,(\textbf{k}_d' - \textbf{p}_d')\, \braket{\hat \chi(\textbf{p},\,t')\,\hat \chi(\textbf{k}-\textbf{p},t')\,\hat \chi(\textbf{p}',\,t'')\,\hat \chi(\textbf{k}'-\textbf{p}',t'')}\,,
\end{align}
that originates from the term proportional to $\chi^4$ in $\langle h_{ij}^{(1)}\,h_{ij}^{(1)}\rangle$.

Finally we have
\begin{align} \label{P01mink}
&\frac{2\,\pi^2}{k^3}\,\delta(\textbf{k}+\textbf{k}')\,{\cal P}_T^{01} =  \frac{1}{8\, \pi^3\, M_P^2} \int dt'\, G_{k'}(t,\,t') \,\Pi_{ij}^{\; \; ab}(\textbf{k}')\, \nonumber \\ 
&\times \int d^3\textbf{p}\,d^3\textbf{p}'\,  \left\langle\hat h^{(0)}_{ij}({\bf k},\,t)\,\hat{h}^{(0)}_{ab}({\bf k}'-{\bf p}-{\bf p}',\,t') \left[\dot\chi(\textbf{p},\,t')\,\dot\chi(\textbf{p}',\,t')+\left({\bf p}\cdot{\bf p}'-m_\chi^2\right)\,\chi(\textbf{p},\,t')\,\chi(\textbf{p}',\,t')\right]\right\rangle\,,
\end{align}
that comes from the cross term between $h_{ij}^{(0)}$ and the part proportional to $h_{ij}^{(0)}$ in eq.~(\ref{solmink}).

Let us now evaluate ${\cal P}_T^{11}$ and ${\cal P}_T^{01}$.

\begin{itemize}

\item ${\cal P}_T^{11}$. To properly calculate particle production in a time-dependent background we decompose $\chi$ using adiabatic mode functions and operators
\begin{align}
&\chi(\textbf{p},t) = \tilde{\chi}_\textbf{p}(t) b(\textbf{p}) + \tilde{\chi}^*_{-\textbf{p}}(t) b^\dagger(-\textbf{p})\,,\nonumber\\
&\tilde{\chi}_\textbf{p}(t) = \frac{1}{\sqrt{2p}}e^{-ipt}\,,\nonumber\\ 
&b(\textbf{p}) = \alpha_p\, a(\textbf{p}) + \beta_p^*\, a^\dagger(-\textbf{p}),
\end{align}
where $\alpha_p$ and $\beta_p$ are the Bogolyubov coefficients which diagonalize the Hamiltonian, and that are evaluated when $\chi$ is massless.  Using Wick's theorem to reduce the four-point function of $\chi$ we find
\begin{align}
&\braket{\chi(\textbf{p},\,t')\,\chi(\textbf{k}-\textbf{p},\,t')\,\chi(\textbf{p}',\,t'')\,\chi(\textbf{k}'-\textbf{p}',\,t'')}\nonumber\\
& = \braket{\chi(\textbf{p},\,t')\,\chi(\textbf{k}-\textbf{p},\,t')} \braket{\chi(\textbf{p}',\,t'')\,\chi(\textbf{k}'-\textbf{p}',\,t'')} \nonumber \\ 
& + \braket{\chi(\textbf{p},\,t')\, \chi(\textbf{p}',\,t'')} \braket{ \chi(\textbf{k}-\textbf{p},\,t')\,\chi(\textbf{k}'-\textbf{p}',\,t'')}\nonumber\\&
 + \braket{\chi(\textbf{p},\,t')\, \chi(\textbf{k}'-\textbf{p}',\,t'')}\braket{\chi(\textbf{k}-\textbf{p},\,t')\, \chi(\textbf{p}',\,t'')}\,.
\end{align}

The first term produces a disconnected term, $\delta^{(3)}(\textbf{k})\, \delta^{(3)}(\textbf{k}')$, which we remove\footnote{We actually want the graviton variance $\sigma^2_h = \braket{(h- \braket{h})\,(h- \braket{h})}$ so that the $\braket{h}\,\braket{h}$ term also produces a disconnected term which cancels with this one.} and the remaining two terms are equivalent.  After normal ordering the $b^{(\dagger)}$ operators, the scalar correlator reads
\begin{equation}  \label{corrbogo}
\braket{ : \chi(\textbf{p},\,t')\chi(\textbf{q},\,t'') : } = \frac{\delta^{(3)}(\textbf{p} + \textbf{q})}{p}\, \left[ |\beta_p|^2\, \cos \left(p\,(t' - t'')\right) + \textrm{Re} \left\{ \alpha_p\, \beta^*_p \,e^{-i\,p\,(t'+ t''))} \right\} \right].
\end{equation}

It is important, and will be relevant for the discussion in Section \ref{sec:finite} in the main text, to note that the prescription~(\ref{corrbogo}) above is equivalent to setting
\begin{equation}  \label{corrsubtr}
\braket{\chi(\textbf{p},\,t')\chi(\textbf{q},\,t'')} = \delta^{(3)}(\textbf{p} + \textbf{q})\, \left[ \chi({\bf p},\,t')\,\chi({\bf p},\,t'')- \tilde\chi({\bf p},\,t')\,\tilde\chi({\bf p},\,t'')\right],
\end{equation}
where $\tilde{\chi}({\bf p},\,t)=\frac{e^{-ipt}}{\sqrt{2\,p}}$ corresponds to the mode function in absence of particle creation, $\Lambda\to 0$.

After performing the momentum summation as in eq.~(\ref{proj}) in the main text, we find
\begin{align}
{\cal P}_T^{11}&= \frac{k}{4 \,\pi^5\, M_P^4}  \int\limits_{0}^t dt' \sin (k (t-t')) \int\limits_{0}^t dt''  \sin (k (t-t'')) \times \nonumber \\ & \times \int \frac{d^3\textbf{p} \; p^4 \sin^4 \theta}{p |\textbf{k} - \textbf{p}|} \left[ |\beta_p|^2 \cos (p(t' - t'')) + \textrm{Re} \left[ \alpha_p \beta^*_p e^{-ip(t'+t''))} \right] \right] \times \nonumber \\ & \times \left[ |\beta_{|\textbf{k}-\textbf{p}|}|^2 \cos (|\textbf{k}-\textbf{p}|(t' - t'')) + \textrm{Re} \left[ \alpha_{|\textbf{k}-\textbf{p}|} \beta^*_{|\textbf{k}-\textbf{p}|} e^{-i|\textbf{k}-\textbf{p}|(t'+ t''))} \right] \right],
\end{align}
where we have used $\textbf{k} \cdot \textbf{p} = k\, p\, \cos \theta$.  We have computed the above integral numerically. Plots of ${\cal P}_T^{11}$ times $M_P^4/k^4$ are shown in Figure \ref{ps11} for $\frac{\Lambda_\chi}{k} = 10, \, 20, \, 30$.  The maximum occurs around $x= kt \approx 2.65$, and a numerical fit shows that the amplitude of the tensor spectrum at its peak goes as
\begin{equation}
\mathcal{P}_T(k) = 1.25 \times 10^{-4} \frac{\Lambda_\chi^5}{M^4_P k} \, , \hspace{1.0cm} x = kt \approx 2.65 \, . 
\end{equation}
The above result shows that ${\cal P}_T^{11}\propto \Lambda_\chi^5/M_P^4$, in accordance with the result obtained on a de Sitter background in the main text.

\begin{figure}
\centering
    \includegraphics[width=0.7\textwidth]{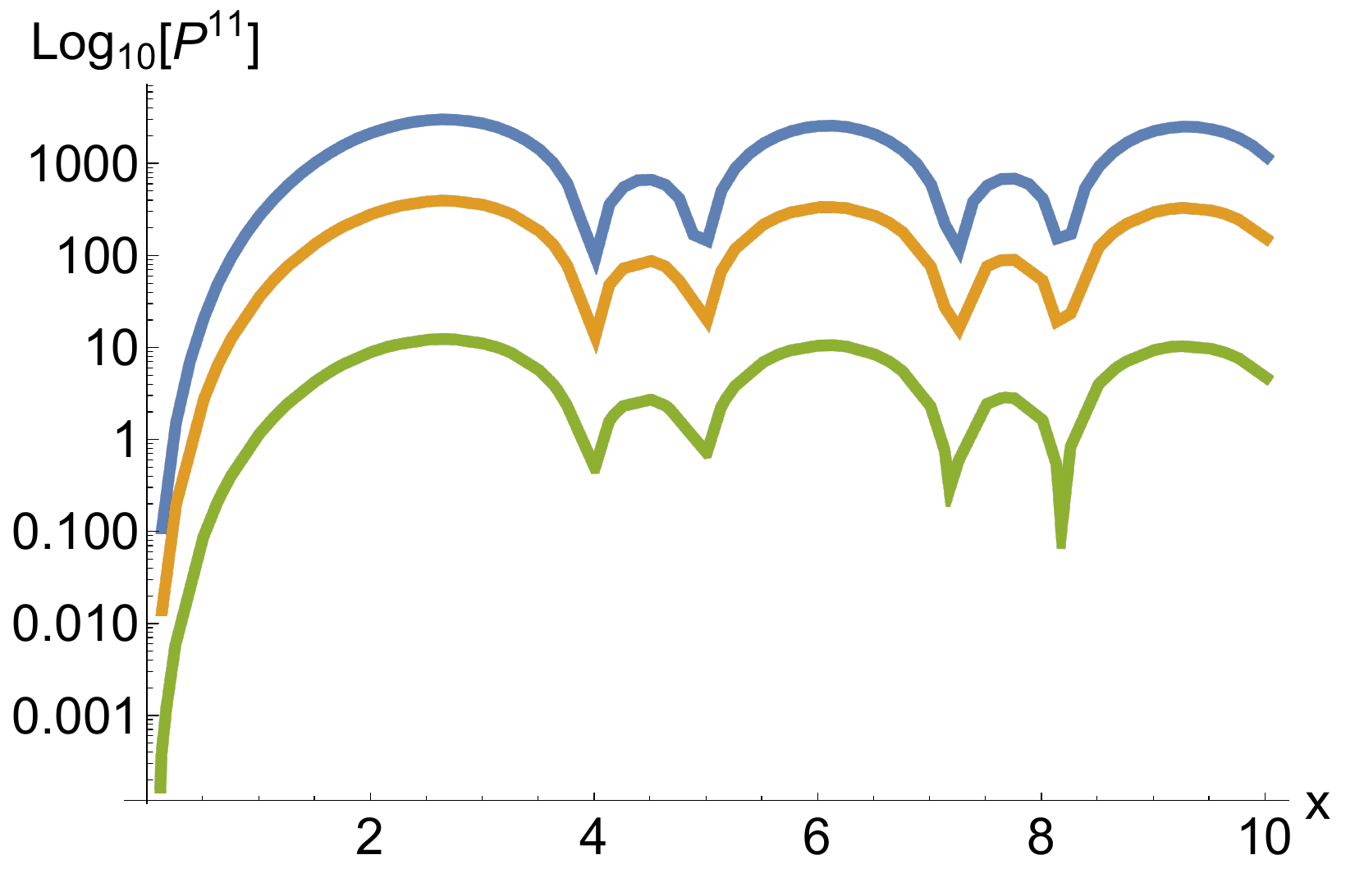}
    \caption{Numerical evaluation of the power spectrum for $\mathcal{P}^{11}$ for $\lambda = 10$ (blue), $\lambda = 20$ (orange), and $\lambda = 30$ (red). The maximum value is attained around $kt \approx 2.65$. }\label{ps11}
\end{figure}

\item ${\cal P}_T^{01}$. Since $h_{ij}^{(0)}$ and $\chi$ are uncorrelated, the expectation value $\langle\dots\rangle$ in eq.~(\ref{P01}) splits into the product $\langle hh\rangle\,\langle \chi\chi\rangle$. To compute the graviton two-point function in momentum space we use the Minkowski decomposition
\begin{equation}
\hat{h}_{ij}^{(0)}(\textbf{p},\,t) = \frac{2}{M_P}\sum_\lambda  \left[v_{\textbf{p}}(t,\,\lambda) \,e_{ij}(\hat{\textbf{p}},\,\lambda)\, \hat{a}_{\textbf{p}}(\lambda)+ v_{-\textbf{p}}^*(t,\,\lambda) \,e_{ij}(\hat{\textbf{p}},\,\lambda)\, \hat{a}_{-\textbf{p}}^\dagger(\lambda) \right]\,,
\end{equation}
where $e_{ij}(\hat{\textbf{p}},\,\lambda)$ is the helicity$-\lambda$ projector. The mode functions $v_{\textbf{p}}(t,\,\lambda)$ in Minkowski space are simply given by the massless plane waves
\begin{align}
v_{\textbf{p}}(t,\,\lambda)=\frac{e^{-ipt}}{\sqrt{2\,p}}\,,
\end{align}
so that
\begin{align}
\langle h_{ij}^{(0)}({\bf k},\,t)\,h_{ab}^{(0)}({\bf k}',\,t')\rangle=\frac{2\,\delta({\bf k}+{\bf k}')}{k\,M_P^2}\,e^{-i\,k\,(t-t')}\sum_\lambda e_{ij}(\hat{\textbf{k}},\,\lambda)\,e_{ab}(\hat{\textbf{k}},\,\lambda)\,,
\end{align}
where
\begin{align}
\sum_\lambda e_{ij}(\hat{\textbf{k}},\,\lambda)\,e_{ab}(\hat{\textbf{k}},\,\lambda)=\Pi_{ij}{}^{ab}({\bf {k}})\, .
\end{align}

In order to proceed we also need the quantity
\begin{equation}  \label{corrbogodot}
\braket{\dot\chi(\textbf{p},\,t')\dot\chi(\textbf{q},\,t'')} = \delta^{(3)}(\textbf{p} + \textbf{q})\,p\,\left[ |\beta_p|^2\, \cos \left(p\,(t' - t'')\right) - \textrm{Re} \left\{ \alpha_p\, \beta^*_p \,e^{-i\,p\,(t'+ t''))} \right\} \right]\,.
\end{equation}
Using the above, we can write the power spectrum as,
\begin{equation}
2 \text{Re}\{\mathcal{P}^{10}_\mathcal{T}\} = - \frac{3 k}{4 \pi^5 M_P^4} \int\limits_0^t dt' \; \sin (2k(t - t')) \int d^3 \textbf{p} \; p \; \textrm{Re}\left[\alpha_p \beta^*_p e^{-2ip t'} \right]  \, .
\end{equation}
Examining the UV behavior of this integral we find that it diverges logarithmically since in the UV the integrand goes as $\frac{dp}{p}$. We integrate up to $p_{UV} = 50 \Lambda_\chi$ to estimate the integral numerically. A plot for $\Lambda_\chi/k = 10, \, 20, \, 30$ is shown in Figure \ref{ps01} and shows that the first peak for all values of $\Lambda_\chi$ occurs around $k\,t\approx 2.37$. Using this value of $k\,t$ we find that the power spectrum at its peak is approximated by
\begin{equation}
2 \text{Re}\{\mathcal{P}^{10}_\mathcal{T}\} = 1.6 \times 10^{-2} \, \, \frac{k \,\Lambda_\chi^3}{M_P^4} \, , \hspace{1.0cm} x = kt \approx 2.37 \, .
\end{equation}
Again, we note that the above result has the same scaling $\propto \Lambda_\chi^3/M_P^4$ as that obtained in the de Sitter calculation of the main text.

\begin{figure}
\centering
    \includegraphics[width=0.7\textwidth]{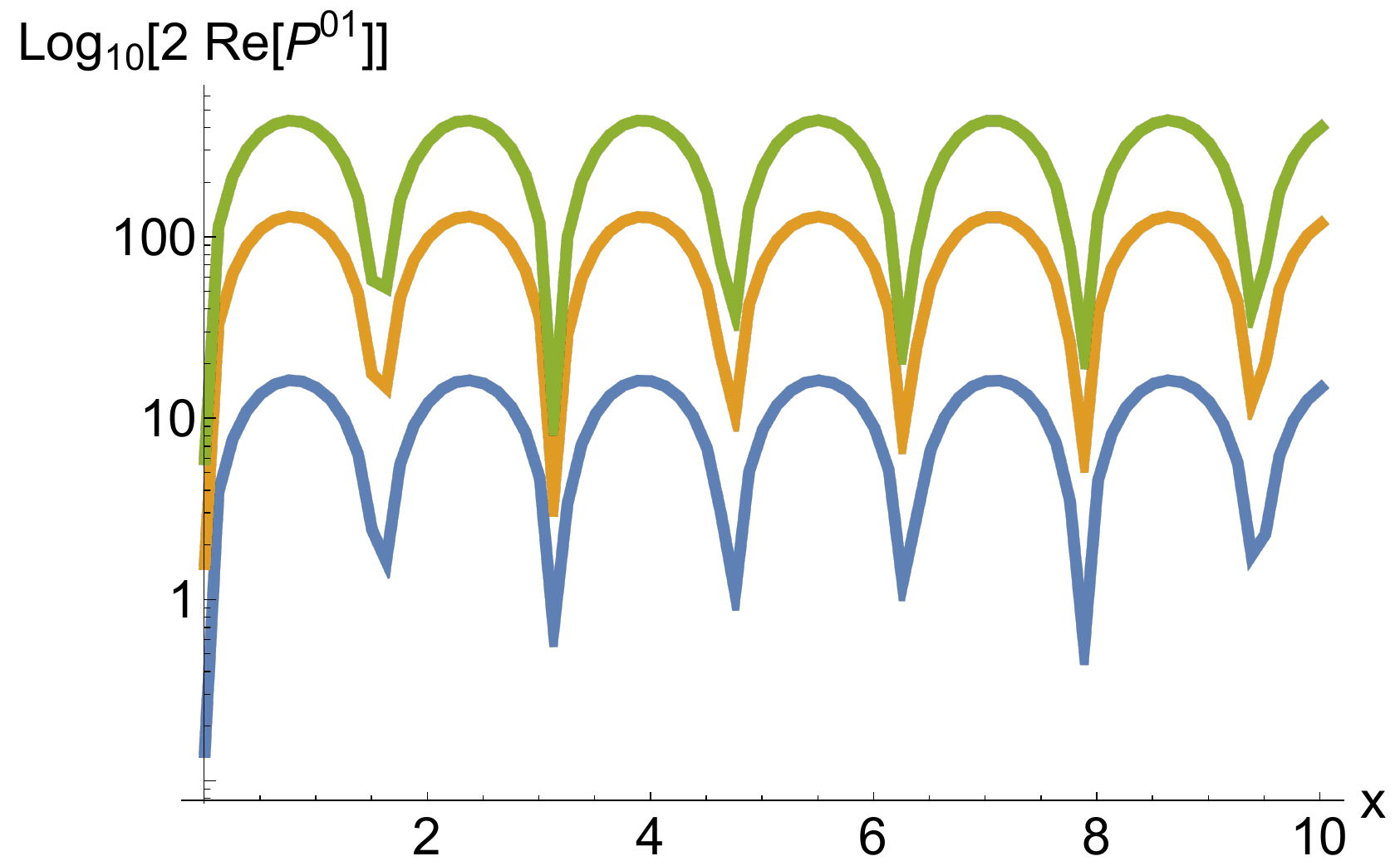}
    \caption{Numerical evaluation of the power spectrum for $2 \text{Re}\{\mathcal{P}^{10}_\mathcal{T}\} $ for $\lambda = 10$ (blue), $\lambda = 20$ (orange), and $\lambda = 30$ (red). The first maximum occurs around $x\approx 2.37$.}\label{ps01}
\end{figure}

\end{itemize}

We finally observe that the two point functions ${\mathcal{P}}^{11}_\mathcal{T}$ and ${\mathcal{P}}^{01}_\mathcal{T}$ oscillate with constant amplitude. This is due to the fact that in Minkowski space the gravitational waves to not dilute away. On a de Sitter background, the amplitude will decrease on timescales of the order of $H^{-1}$.


\end{document}